\documentclass[prd,aps,floatfix,nofootinbib, 11 pt]{revtex4}

\usepackage{amsmath,graphicx,color,epsfig}

\begin{document}

\title{ Rare decays of $\Lambda_b \to \Lambda \gamma$ and $\Lambda_b \to \Lambda l^{+} l^{-}$
in universal extra dimension model}
\author{Yu-Ming Wang$^{1}$}
\author{M. Jamil Aslam$^{1,2}$}
\author{Cai-Dian L\"{u} $^{1}$}

\affiliation{$^{1}$Institute of High Energy Physics, P.O. Box
918(4), Beijing 100049, China} \affiliation{$^{2}$National Center
for Physics, Quaid-i-Azam University, Islamabad, Pakistan}

\begin{abstract}
The exclusive weak decay of $\Lambda_b \to \Lambda \gamma$ and
$\Lambda_b \to \Lambda l^{+} l^{-}$ are investigated in the
Applequist-Cheng-Dobrescu  model, which is an extension of the
standard model in presence of universal extra dimensions. Employing
the transition form factors obtained in the light-cone sum rues, we
analyze how the invariant mass distribution, forward-backward
asymmetry and polarization asymmetry of $\Lambda$ baryon of these
decay modes can be used to constrain the only one additional free
parameter with respect to the standard model, namely, the radius $R$
of the extra dimension. Our results indicate that the Kaluza-Klein
modes can lead to approximately 25\% suppression of the branching
ratio of $\Lambda_b \to \Lambda \gamma$, however, their
contributions can bring about 10\% enhancement to the decay rate of
$\Lambda_b \to \Lambda l^{+} l^{-}$. It is shown that the
zero-position of forward-backward asymmetry of $\Lambda_b \to
\Lambda \mu^{+} \mu^{-}$ is sensitive to the compactification
parameter $R$ in this scenario, while the measurement of
polarizations of $\Lambda$ baryon in the $\Lambda_b$ decays is not
suitable to provide some valuable information for the universal
extra dimension model.

\end{abstract}

\maketitle

\section{Introduction}

At present, the Standard Model (SM) of particle physics has been
scrutinized relentlessly from its inception and held its ground over
the entire breadth of its theoretical reach almost without failure.
In spite of its impressive successes, the SM is not completely
satisfactory as the theory of elementary particles from the view
point of both aesthetics and phenomenology. It has been  realized
that bottom quark physics is a powerful probe of physics beyond the
SM in a  complementary way to the direct searches, which is crucial
to identify the new physics (NP) and its properties correctly as
well as understand its theoretical consequences. Rare decays
involving $b \to s$ flavor changing neutral current (FCNC), which
are forbidden at the tree level in the SM, can provide an ideal
platform to test the SM precisely as well as bound its extensions
stringently so that pave the way for the establishment of new
physics beyond the SM. Wealthy experimental data on both inclusive
and exclusive $b \to s$ FCNC $B$ meson decays \cite{HFAG} have been
accumulated at the $e^{+} e^{-}$ factories operating at the peak of
$\Upsilon(4S)$, which also motivated intensive theoretical studies
on these mesonic decay modes.

Unlike mesonic decays, the investigations of FCNC $b \to s$
transition for bottom baryonic decays $\Lambda_b \to \Lambda \gamma$
and $\Lambda_b \to \Lambda l^{+} l^{-}$
\cite{Mannel,Mohanta,Cheng,Cheng 2,Huang,Hiller:2001zj,
HLLW,Aslam:2008hp} are much behind because more degrees of freedom
are involved in the bound state of baryon system at the quark level.
It should be pointed out that such baryonic decays can offer the
unique ground to extract the helicity structure of effective
Hamiltonian for FCNC $b \to s$ transition in the SM and beyond,
since the information on the handness of the quark is lost in the
hadronization of meson case. Compared with the $B$ meson decays,
$\Lambda_b$ baryon decays contain some particular observables
involving the spin of $b$ quark, which are sensitive to the new
physics and more easily detectable. From the viewpoint of
experiment, the only drawback of bottom baryon decays is that the
production rate of $\Lambda_b$ baryon in $b$ quark hadronization is
about four times less than that of  $B$ meson, hence we need more
experimental data on heavy quark decays from future colliders, such
as Large Hadron Colliders (LHC) at CERN, and Tevatron collider at
FNAL, to perform a stringent constraint on the parameter space of
available new physics models.

Among various models of physics beyond the SM, the models with extra
dimensions \cite{Antoniadis:1990ew} are of intensive interest, since
they provide a unified framework for gravity and other interactions,
which can give some hints of the hierarchy problem and a connection
with string theory. Among the extra dimension models, the models of
particular interests are the scenarios with universal extra
dimensions (UED), which are the most democratic extra dimension
model in the scene that all SM fields are allowed to propagate in
the extra dimension. Above the compactification scale $1/R$, a given
UED model becomes a higher dimensional field theory whose equivalent
description in four dimensions consists of the SM fields, the tower
of their Kaluza-Klein (KK) partners and additional tower of KK modes
without corresponding SM partners. A simple scenario is represented
by the Applequist-Cheng-Dobrescu (ACD) model \cite{ACD} with a
single compactified extra dimension, which introduces  only one
additional free parameter relative to the SM, i.e. $1/R$, the
inverse of the compactification radius. As all particles can access
the bulk, momentum along the fifth dimension, and hence the KK
number is conserved in any process, which can break down to the
conservation of KK-parity, defined as $(-1)^n$ due to the orbifold
corrections.

Different bounds to the size of extra dimension have been explored
in various processes already accessible at particle accelators or
within the reach of future facilities. The analysis on the Tevatron
run I data allows to establish the bound $1/R \geq 250-300 \rm{GeV}$
\cite{Appelquist:2002wb}. Analysis of the anomalous magnetic moment
\cite{Agashe:2001ra, Appelquist:2001jz} and the $Z \to b \bar{b}$
vertex \cite{Oliver:2002up} also gives rise to the bound $1/R \geq
300 \rm{GeV}$. The conservation of KK parity implies the absence of
tree level KK contributions to low energy processes taking place at
scales $\mu \ll 1/R$. The fact that KK excitations  can influence
processes occurring at loop level also indicates that $b \to s$ FCNC
transitions are extremely important for constraining the extra
dimension model.  For this reason, the effective Hamiltonian
responsible for $b \to s$ transitions was derived in
\cite{Buras:2002ej, Buras:2003mk}.  $K_L-K_S$ mass difference, the
parameter $\epsilon_K$, $B^0_{d,s}-\bar{B}^0_{d,s}$ mixing mass
differences $\Delta M_{d,s}$, rare decays of $K$ and $B$ meson as
well as CP-violating ratio $\epsilon^{\prime}/\epsilon$ are also
comprehensively studied there. In particular, it is found that $BR(B
\to X_s \gamma)$ allows to constrain $1/R \geq 250 \rm{GeV}$
\cite{Buras:2003mk}, which has been updated by a more recent
analysis in combination with the NNLO values of  Wilson coefficient
in the SM and new experimental data to $1/R \geq 600 \rm{GeV}$ at
95\% C.L. \cite{Haisch:2007vb}. Exclusive  $B \to K^{\ast}(K_{1})
l^{+} l^{-}$, $B \to K^{\ast} \nu \bar{\nu}$ and $ B \to K^{\ast}
\gamma$ decays of $B$ meson \cite{Colangelo} together with  $B_s \to
\phi l^{+} l^{-}$ and  $B_s \to \gamma l^{+} l^{-}$ decays of $B_s$
meson \cite{Mohanta:2006ae} are also studied in the framework of the
UED scenario.

Moreover, $\Lambda_b \to \Lambda \gamma$ and $\Lambda_b \to \nu
\bar{\nu}$ decays in the UED model are further  investigated in
\cite{Colangelo:2007jy} employing the form factors calculated in the
three-point QCD sum rules (QCDSR) within the framework of heavy
quark effective theory (HQET) \cite{Huang}. The sensitivity of
branching ratio, forward-backward asymmetry and polarization
asymmetry of lepton for semileptonic decay of $\Lambda_b \to \Lambda
l^{+} l^{-}$ to the compactification parameter $1/R$ are analyzed in
\cite{Aliev:2006xd,Aliev:2006gv} using the transition form factors
given by the QCDSR at length. In this work, we would like to revisit
$\Lambda_b \to \Lambda \gamma$  and $\Lambda_b \to \Lambda l^{+}
l^{-}$ decays in the ACD model with the form factors derived in the
light-cone sum rules (LCSR) \cite{Wang:2008sm}, where the effects of
higher twist distribution amplitudes of $\Lambda$ baryon are
included. In particular, we consider how  the polarization asymmetry
of $\Lambda$ baryon in $\Lambda_b \to \Lambda l^{+} l^{-}$ decays
can be used to constrain the radius of extra dimension, which are
still not available in the literature. The structure of this paper
is organized as follows: After a brief introduction to the ACD model
in section II, we present the effective Hamiltonian for $b \to s$
transition and parameterizations of transition form factors in
section III. The dependence of branching ratio, forward-backward
asymmetry and polarization asymmetry of $\Lambda$ baryon  in
$\Lambda_b \to \Lambda l^{+} l^{-}$ decays on the radius of extra
dimension $R$ are given in section IV, where the sensitivity of
branching fraction of  $\Lambda_b \to \Lambda \gamma$ on the
compactification parameter $R$ is also presented. The last section
is devoted to the conclusion.

\section{Review of ACD model}

In our usual universe we have 3 spatial plus 1 temporal dimensions
and if an extra dimension exists and is compactified, fields living
in all dimensions would manifest themselves in the $3+1$ space by
the appearance of KK excitations. The most pertinent question is
whether ordinary fields propagate or not in all extra dimensions.
One obvious possibility is the propagation of gravity in whole
ordinary plus extra dimensional universe, the ``bulk''. Contrary to
this there are models with UED in which all the fields propagate in
all available dimensions \cite{ACD} and ACD model belongs to one of
UED scenarios \cite{Colangelo}

This model is the minimal extension of the SM in $4+\delta $
dimensions, and in literature a simple case $\delta =1$ is
considered \cite{Colangelo}. The topology for this extra dimension
is orbifold $S^{1}/Z_{2}$, and the coordinate $x_{5}=y$ runs from
$0$ to $2\pi R$, where $R$ is the compactification radius. The KK
mode expansion of the fields
are determined from the boundary conditions at two fixed points $y=0$ and $%
y=\pi R$ on the orbifold. Under parity transformation $P_{5}:$
$y\rightarrow -y$ the fields may be even or odd. Even fields have
their correspondent in the $4$-dimensional SM and their zero mode in
the KK mode expansion can be interpreted as the ordinary SM field.
The odd fields do not have their correspondent in the SM and
therefore do not have zero mode in the KK expansion.

The salient features of the ACD model are:

\begin{itemize}

\item the compactification radius $R$ is the only free parameter with
respect to SM

\item no tree level contribution of KK modes in low energy processes
(at scale $\mu \ll 1/R$) and no production of single KK excitation
in ordinary particle interactions are the consequences of the
conservation of KK parity.
\end{itemize}

The detailed description of ACD model is provided in
\cite{Buras:2002ej}; here we summarize main features of its
construction from the Ref. \cite{Colangelo}.

\textbf{Gauge group}

As ACD model is the minimal extension of SM, therefore the gauge
bosons associated with the gauge group $SU\left( 2\right) _{L}\times
U\left( 1\right) _{Y}$ are $W_{i}^{a}\,(a=1,\,2,\,3$,
$i=0$,$\,1$,$\,2$,$\,3$,$\,5)$ and $B_{i}$. The gauge couplings are
$\hat{g}_{2}=g_{2}\sqrt{2\pi R}$ and $\hat{g}^{\prime }=g^{\prime
}\sqrt{2\pi R}$ (the hat on the coupling constant refers to the
extra dimension). The charged bosons are $W_{i}^{\pm
}=\frac{1}{\sqrt{2}}\left( W_{i}^{1}\mp W_{i}^{2}\right) $ and the
mixing of $W_{i}^{3}$ and $B_{i}$ gives rise to the fields $Z_{i}$
and $A_{i}$ as they do in the SM. The relations for the mixing
angles are:
\begin{equation}
c_{W}=\cos \theta _{W}=\frac{\hat{g}_{2}}{\sqrt{\hat{g}_{2}^{2}+\hat{g}%
^{\prime 2}}},  \qquad s_{W}=\sin \theta _{W}=\frac{\hat{g}^{\prime
}}{\sqrt{\hat{g}_{2}^{2}+\hat{g}^{\prime 2}}}. \label{couplings}
\end{equation}
The Weinberg angle remains the same as in the SM, due to the
relationship between five and four dimensional constants. The gluons
which are the gauge bosons associated to $SU\left( 3\right) _{C}$
are $G_{i}^{a}\left( x,y\right) (a=1,\ldots ,8)$.

\textbf{Higgs sector and mixing between Higgs fields and gauge
bosons}

The Higgs doublet can be written as:
\begin{equation}
\phi =\left(
\begin{array}{l}
\hspace {0.8 cm} i\chi ^{+} \\
\frac{1}{\sqrt{2}}\left( \psi -i\chi ^{3}\right)%
\end{array}
\right)  \label{higgs-doublet}
\end{equation}
with $\chi ^{\pm }=\frac{1}{\sqrt{2}}\left( \chi ^{1}\mp \chi
^{2}\right) $. Now only field $\psi $ has a zero mode, and we assign
vacuum expectation value $\hat{v}$ to such mode, so that $\psi
\rightarrow \hat{v}+H$. $H$ is the SM Higgs field, and the relation
between expectation values in five and four dimension is:
$\hat{v}=v/\sqrt{2\pi R}$.

The Goldstone fields $G_{\left( n\right) }^{0}$, $G_{\left( n\right)
}^{\pm} $ arise due to the mixing of charged $W_{5\left( n\right) }^{\pm }$ and $%
\chi _{\left( n\right) }^{\pm }$ , as well as neutral fields
$Z_{5\left(
n\right) }$. These Goldstone modes are then used to give masses to the $%
W_{\left( n\right) }^{\pm \mu }$ and $Z_{\left( n\right) }^{\mu }$, and $%
a_{\left( n\right) }^{0}$, $a_{\left( n\right) }^{\pm }$, new
physical scalars.

\textbf{Yukawa terms}

In SM, Yukawa coupling of the Higgs field to the fermion provides
the fermion mass terms. The diagonalization of such terms leads to
the introduction of the CKM matrix. In order to have chiral fermions
in ACD model, the left and right-handed components of the given
spinor cannot be simultaneously even under $P_{5}$.\ This makes the
ACD model to be the minimal flavor violation model, since there are
no new operators beyond those present in the SM and no new phases
beyond the CKM phase. The unitarity triangle remains the same as in\
SM \cite{Buras:2002ej}. In order to have 4D mass eigenstates of
higher KK levels, a further mixing is introduced among the
left-handed doublet and right-handed singlet of each flavor $f$. The
mixing
angle is  $\tan \left( 2\alpha _{f\left( n\right) }\right) =\frac{%
m_{f}}{n/R}\left( n\geq 1\right) $ giving mass $m_{f\left( n\right) }=\sqrt{%
m_{f}^{2}+\frac{n^{2}}{R^{2}}}$, so that it is negligible for all
flavors except the top \cite{Colangelo}.

Integrating over the fifth-dimension $y$, one can  gain the
four-dimensional Lagrangian
\begin{equation}
\mathcal{L}_{4}\left( x\right) =\int_{0}^{2\pi
R}\mathcal{L}_{5}\left( x,y\right) dy \label{4-dllagrangian}
\end{equation}
which describes zero modes corresponding to the SM  fields and their
massive KK excitations together with KK excitations without zero
modes which do not corresponds to any field in SM. The Feynman rules
used in the further calculation are given in Ref.
\cite{Buras:2002ej}.

\section{Effective Hamiltonian and transition form factors}

\subsection{Effective Hamiltonian }

Integrating out the particles including top quark, $W^{\pm}$ and $Z$
bosons above scale $\mu=O(m_b)$ , we  arrive at the effective
Hamiltonian responsible for the $b \to s l^{+}l^{-}$ transition in
the SM \cite{Buchalla:1995vs}
\begin{eqnarray}
H_{eff}(b\to s l^+l^-) &=& -\frac{G_{F}}{2\sqrt{2}}V_{tb}V_{ts}^{*}
\bigg[ {\sum\limits_{i=1}^{6}} C_{i}({\mu}) Q_{i}({\mu})+C_{7
\gamma}(\mu) Q_{7\gamma}(\mu)+C_{8G}(\mu) Q_{8G}(\mu)\nonumber\\
&&+C_{9}(\mu) Q_{9 }(\mu)+C_{10}(\mu) Q_{10}(\mu) \bigg],
\label{effective haniltonian 1}
\end{eqnarray}
where we have neglected the terms proportional to $V_{ub}V_{us}^{*}$
on account of $|V_{ub}V_{us}^{*}/V_{tb}V_{ts}^{*}|<0.02$. The
complete list of the operators can be given by
\begin{itemize}
 \item  current--current (tree) operators
    \begin{eqnarray}
  Q^{u}_{1}=({\bar{s}}_{\alpha}c_{\beta} )_{V-A}
               ({\bar{c}}_{\beta} b_{\alpha})_{V-A},
    \ \ \ \ \ \ \ \ \
   Q^{u}_{2}=({\bar{s}}_{\alpha}c_{\alpha})_{V-A}
               ({\bar{c}}_{\beta} b_{\beta} )_{V-A},
    \label{eq:operator02}
    \end{eqnarray}
     \item  QCD penguin operators
    \begin{eqnarray}
      Q_{3}=({\bar{s}}_{\alpha}b_{\alpha})_{V-A}\sum\limits_{q}
           ({\bar{q}}_{\beta} q_{\beta} )_{V-A},
    \ \ \ \ \ \ \ \ \
    Q_{4}=({\bar{s}}_{\beta}b_{\alpha})_{V-A}\sum\limits_{q}
           ({\bar{q}}_{\alpha} q_{\beta} )_{V-A},
    \label{eq:operator34} \\
     \!\!\!\! \!\!\!\! \!\!\!\! \!\!\!\! \!\!\!\! \!\!\!\!
    Q_{5}=({\bar{s}}_{\alpha}b_{\alpha})_{V-A}\sum\limits_{q}
           ({\bar{q}}_{\beta} q_{\beta} )_{V+A},
    \ \ \ \ \ \ \ \ \
    Q_{6}=({\bar{s}}_{\beta}b_{\alpha})_{V-A}\sum\limits_{q}
           ({\bar{q}}_{\alpha} q_{\beta} )_{V+A},
    \label{eq:operator56}
    \end{eqnarray}
 \item magnetic penguin operators
    \begin{eqnarray}
     Q_{7 \gamma}={e \over 8\pi^2} \bar{s}_{\alpha}\sigma^{\mu \nu}(m_b R+m_sL)b_{\alpha}\, F_{\mu \nu},
    \ \ \ \
    Q_{8 G}={g \over 8\pi^2} \bar{s}_{\alpha}\sigma^{\mu \nu}(m_b R+m_sL)T^a_{\alpha \beta}b_{\beta}\,
    G^{a}_{\mu \nu },
    \label{eq:operator78}
        \end{eqnarray}
     \item semi-leptonic operators
    \begin{eqnarray}
     Q_{9V}= ({\bar{s}}_{\alpha}b_{\alpha})_{V-A}({\bar{e}}e)_{V},
    \ \ \ \
   Q_{10A}= ({\bar{s}}_{\alpha}b_{\alpha})_{V-A}({\bar{e}}e)_{A},
    \label{eq:operator910}
    \end{eqnarray}
 \end{itemize}
where $\alpha$ and $\beta$ are the color indices, $R(L)=1 \pm
\gamma_5, \sigma_{\mu\nu}={i\over 2}[\gamma_\mu,\gamma_\nu]$, $e$
and $g$ are the coupling constant of electromagnetic and strong
interactions, respectively; and $q$ are the active quarks at the
scale $\mu=O(m_b)$, i.e.
 $q=(u,d,s,c,b)$.
The left handed current is defined as $({\bar{q}}_{\alpha} q_{\beta}
)_{V - A}= {\bar{q}}_{\alpha} \gamma_\nu (1 - \gamma_5) q_{\beta} $
and the corresponding right handed current is $({\bar{q}}_{\alpha}
q_{\beta} )_{V+A}= {\bar{q}}_{\alpha} \gamma_\nu (1+\gamma_5)
q_{\beta}  $.

In terms of the Hamiltonian given in Eq. (\ref{effective haniltonian
1}), we can derive the free quark decay amplitude for $b \to s
l^{+}l^{-}$ process as
\begin{eqnarray}
M(b\to s l^+l^-) &=& \frac{G_{F}}{2\sqrt{2}}V_{tb}V_{ts}^{*}
{\alpha_{em}\over \pi}\bigg \{-{2i \over q^2}C_7 (\mu) \bar{s}
\sigma_{\mu\nu}q^{\nu}(m_b R+m_sL) b  \bar{l}\gamma^{\mu}l \nonumber
\\&&+ C_{9}(\mu)\bar{s}\gamma_{\mu}Lb
\bar{l}\gamma^{\mu}l+C_{10}
\bar{s}\gamma_{\mu}Lb\bar{l}\gamma^{\mu}\gamma_5l \bigg\}. \label{b
to s l l}
\end{eqnarray}
Similarly, the free quark decay amplitude for $b\to s\gamma$ can be
written as
\begin{eqnarray}
M(b\to s\gamma) &=& \frac{G_{F}}{2\sqrt{2}}V_{tb}V_{ts}^{*} {e\over
4\pi^2} C_7(\mu) \bar{s} \sigma_{\mu\nu}q^{\nu}(m_b R+m_sL) b
F^{\mu\nu}. \label{b to s gamma}
\end{eqnarray}

No operators other than those collected in Eq. (\ref{effective
haniltonian 1}) are found in the ACD model, therefore the effects of
KK contributions are implemented by modifying the Wilson
coefficients that now depend on the additional ACD parameter, the
compactification radius; if we neglect the contributions of scalar
fields, which are indeed very small. Since the KK states become more
and more massive with large value of $1/R$, which can decouple from
the low-energy theory, hence the SM phenomenology should be
recovered in the limit $1/R \to +\infty$. It also needs to emphasize
that we do not include the long-distance contributions from
four-quark operators near the $c\bar{c}$ resonance, which can be
experimentally removed applying appropriate kinematical cuts in the
neighborhood of resonance region. Besides, the QCD penguin operators
are also neglected due to their small Wilson coefficients compared
to the others. Therefore, we only need to specify the Wilson
coefficients $C_7$, $C_9$ and $C_{10}$, which have been given in
\cite{Buras:2003mk}.  It is found that the impact of the KK states
results in the enhancement of $C_{10}$ and the suppression of $C_7$.

As a general expression, the Wilson coefficients are represented by
functions $F(x_t, 1/R)$ generalizing the SM analogues $F_0(x_t)$:
\begin{eqnarray}
F(x_t. 1/R) =F_0(x_t) + \sum_{n=1}^{\infty} F_n(x_t, x_n),
\label{general function}
\end{eqnarray}
where $x_n={ m_n^2 \over m_W^2}$ and $m_n={n \over R}$. A remarkable
feature is that the sum over the KK contributions in Eq.
(\ref{general function}) is finite at leading order in all cases as
a result of a generalized GIM mechanism \cite{Buras:2002ej}. The
relevant functions are the following: $C(x_t,1/R)$  from $Z^0$
penguins; $D(x_t,1/R)$ from $\gamma$ penguins; $E(x_t,1/R)$ from
gluon penguins; $D^\prime (x_t,1/R)$ from $\gamma$ magnetic
penguins; $E^\prime (x_t,1/R)$ from chromomagnetic penguins. They
can  be found in \cite{Buras:2002ej,Buras:2003mk,Colangelo} and are
collected as below.

$\bullet C_{7}$

Here one defines an effective coefficient $C_{7}^{(0)eff}$ which is
renormalization scheme independent \cite{Buras:1993xp}:
\begin{equation}
C_{7}^{(0)eff}(\mu _{b})=\eta ^{\frac{16}{23}}C_{7}^{(0)}(\mu _{w})+\frac{8}{%
3}(\eta ^{\frac{14}{23}}-\eta ^{\frac{16}{23}})C_{8}^{(0)}(\mu
_{w})+C_{2}^{(0)}(\mu _{w})\sum_{i=1}^{8}h_{i}\eta ^{\alpha _{i}},
\label{wilson1}
\end{equation}
where $\eta =\frac{\alpha_s(\mu _{w})}{\alpha _{s}(\mu _{b})},$ and
\begin{equation}
C_{2}^{(0)}(\mu _{w})=1,\text{ }C_{7}^{(0)}(\mu
_{w})=-\frac{1}{2}D^{\prime }(x_{t},\frac{1}{R}),\text{
}C_{8}^{(0)}(\mu _{w})=-\frac{1}{2}E^{\prime }(x_{t},\frac{1}{R});
\label{wilson2}
\end{equation}
the superscript $(0)$ stays for leading logarithm approximation. The
involved parameters in Eq. (\ref{wilson1}) are grouped as
\begin{eqnarray}
\alpha _{1} &=&\frac{14}{23}\text{ \quad }\alpha
_{2}=\frac{16}{23}\text{ \quad }\alpha _{3}=\frac{6}{23}\text{ \quad
}\alpha _{4}=-\frac{12}{23}
\nonumber \\
\alpha _{5} &=&0.4086\text{ \quad }\alpha _{6}=-0.4230\text{ \quad
}\alpha
_{7}=-0.8994\text{ \quad }\alpha _{8}=-0.1456  \nonumber \\
h_{1} &=&2.996\text{ \quad }h_{2}=-1.0880\text{ \quad }h_{3}=-\frac{3}{7}%
\text{ \quad }h_{4}=-\frac{1}{14}  \nonumber \\
h_{5} &=&-0.649\text{ \quad }h_{6}=-0.0380\text{ \quad
}h_{7}=-0.0185\text{ \quad }h_{8}=-0.0057.  \label{wilson3}
\end{eqnarray}
The functions $D^{\prime }$ and $E^{\prime }$ are determined by  Eq. (\ref{wilson3}%
) with
\begin{equation}
D_{0}^{\prime }(x_{t})=-\frac{(8x_{t}^{3}+5x_{t}^{2}-7x_{t})}{12(1-x_{t})^{3}%
}+\frac{x_{t}^{2}(2-3x_{t})}{2(1-x_{t})^{4}}\ln x_{t},
\label{wilson4}
\end{equation}
\begin{equation}
E_{0}^{\prime }(x_{t})=-\frac{x_{t}(x_{t}^{2}-5x_{t}-2)}{4(1-x_{t})^{3}}+%
\frac{3x_{t}^{2}}{2(1-x_{t})^{4}}\ln x_{t},  \label{wilson5}
\end{equation}
\begin{eqnarray}
D_{n}^{\prime }(x_{t},x_{n}) &=&\frac{%
x_{t}(-37+44x_{t}+17x_{t}^{2}+6x_{n}^{2}(10-9x_{t}+3x_{t}^{2})-3x_{n}(21-54x_{t}+17x_{t}^{2}))%
}{36(x_{t}-1)^{3}}  \nonumber \\
&&+\frac{x_{n}(2-7x_{n}+3x_{n}^{2})}{6}\ln \frac{x_{n}}{1+x_{n}}
\nonumber
\\
&&-\frac{%
(-2+x_{n}+3x_{t})(x_{t}+3x_{t}^{2}+x_{n}^{2}(3+x_{t})-x_{n})(1+(-10+x_{t})x_{t}))%
}{6(x_{t}-1)^{4}}  \ln \frac{x_{n}+x_{t}}{1+x_{n}}, \nonumber
\\ \label{wilson6}
\end{eqnarray}
\begin{eqnarray}
E_{n}^{\prime }(x_{t},x_{n}) &=&\frac{%
x_{t}(-17-8x_{t}+x_{t}^{2}+3x_{n}(21-6x_{t}+x_{t}^{2})-6x_{n}^{2}(10-9x_{t}+3x_{t}^{2}))%
}{12(x_{t}-1)^{3}}  \nonumber \\
&&+-\frac{1}{2}x_{n}(1+x_{n})(-1+3x_{n})\ln \frac{x_{n}}{1+x_{n}}
\nonumber
\\
&&+\frac{%
(1+x_{n})(x_{t}+3x_{t}^{2}+x_{n}^{2}(3+x_{t})-x_{n}(1+(-10+x_{t})x_{t}))}{%
2(x_{t}-1)^{4}}\ln \frac{x_{n}+x_{t}}{1+x_{n}}.  \label{wilson7}
\end{eqnarray}
Following \cite{Buras:2002ej},  one can obtain the expressions for
the sum over $n$ as
\begin{eqnarray}
\sum_{n=1}^{\infty }D_{n}^{\prime }(x_{t},x_{n}) &=&-\frac{%
x_{t}(-37+x_{t}(44+17x_{t}))}{72(x_{t}-1)^{3}}  \nonumber \\
&&+\frac{\pi M_{w}R}{2}[\int_{0}^{1}dy\frac{2y^{\frac{1}{2}}+7y^{\frac{3}{2}%
}+3y^{\frac{5}{2}}}{6}]\coth (\pi M_{w}R\sqrt{y})  \nonumber \\
&&+\frac{(-2+x_{t})x_{t}(1+3x_{t})}{6(x_{t}-1)^{4}}J(R,-\frac{1}{2})
\nonumber \\
&&-\frac{1}{6(x_{t}-1)^{4}}%
[x_{t}(1+3x_{t})-(-2+3x_{t})(1+(-10+x_{t})x_{t})]J(R,\frac{1}{2})
\nonumber
\\
&&+\frac{1}{6(x_{t}-1)^{4}}[(-2+3x_{t})(3+x_{t})-(1+(-10+x_{t})x_{t})]J(R,%
\frac{3}{2})  \nonumber \\
&&-\frac{(3+x_{t})}{6(x_{t}-1)^{4}}J(R,\frac{5}{2})],
\label{wilson8}
\end{eqnarray}
\begin{eqnarray}
\sum_{n=1}^{\infty }E_{n}^{\prime }(x_{t},x_{n}) &=&-\frac{%
x_{t}(-17+(-8+x_{t})x_{t})}{24(x_{t}-1)^{3}}  \nonumber \\
&&+\frac{\pi M_{w}R}{2}[\int_{0}^{1}dy(y^{\frac{1}{2}}+2y^{\frac{3}{2}}-3y^{%
\frac{5}{2}})\coth (\pi M_{w}R\sqrt{y})]  \nonumber \\
&&-\frac{x_{t}(1+3x_{t})}{(x_{t}-1)^{4}}J(R,-\frac{1}{2})  \nonumber \\
&&+\frac{1}{(x_{t}-1)^{4}}[x_{t}(1+3x_{t})-(1+(-10+x_{t})x_{t})]J(R,\frac{1}{%
2})  \nonumber \\
&&-\frac{1}{(x_{t}-1)^{4}}[(3+x_{t})-(1+(-10+x_{t})x_{t})]J(R,\frac{3}{2})
\nonumber \\
&&+\frac{(3+x_{t})}{(x_{t}-1)^{4}}J(R,\frac{5}{2})],
\label{wilson9}
\end{eqnarray}
where
\begin{equation}
J(R,\alpha )=\int_{0}^{1}dyy^{\alpha }[\coth (\pi M_{w}R\sqrt{y}%
)-x_{t}^{1+\alpha }\coth (\pi m_{t}R\sqrt{y})].  \label{wilson10}
\end{equation}
$\bullet C_{9}$

In the ACD model and in the naive dimension regularization (NDR)
scheme, the Wilson coefficient $C_9$ can be written as
\begin{equation}
C_{9}(\mu )=P_{0}^{NDR}+\frac{Y(x_{t},\frac{1}{R})}{\sin ^{2}\theta _{W}}%
-4Z(x_{t},\frac{1}{R})+P_{E}E(x_{t},\frac{1}{R}),  \label{wilson11}
\end{equation}
where $P_{0}^{NDR}=2.60\pm 0.25$ \cite{Misiak:1992bc,Buras:1994dj}
and the last term is numerically negligible. The function $Y$ and
$Z$ are given by
\begin{eqnarray}
Y(x_{t},\frac{1}{R}) &=&Y_{0}(x_{t})+\sum_{n=1}^{\infty
}C_{n}(x_{t},x_{n}),
\nonumber \\
Z(x_{t},\frac{1}{R}) &=&Z_{0}(x_{t})+\sum_{n=1}^{\infty
}C_{n}(x_{t},x_{n}),  \label{wilson12}
\end{eqnarray}
where $Y_0(x_t)$, $Z_0(x_t)$ and $C_n(x_t, x_n)$ are
\begin{eqnarray}
\label{wilson13}
Y_{0}(x_{t}) &=&\frac{x_{t}}{8}[\frac{x_{t}-4}{x_{t}-1}+\frac{3x_{t}}{%
(x_{t}-1)^{2}}\ln x_{t}] , \nonumber \\
Z_{0}(x_{t}) &=&\frac{18x_{t}^{4}-163x_{t}^{3}+259x_{t}^{2}-108x_{t}}{%
144(x_{t}-1)^{3}}  \nonumber \\
&&+[\frac{32x_{t}^{4}-38x_{t}^{3}+15x_{t}^{2}-18x_{t}}{72(x_{t}-1)^{4}}-%
\frac{1}{9}]\ln x_{t}, \nonumber \\
C_{n}(x_{t},x_{n})&=&\frac{x_{t}}{8(x_{t}-1)^{2}}%
[x_{t}^{2}-8x_{t}+7+(3+3x_{t}+7x_{n}-x_{t}x_{n})\ln \frac{x_{t}+x_{n}}{%
1+x_{n}}].  \label{wilson14}
\end{eqnarray}
The sum of $C_n(x_t, x_n)$ over $n$ is computed as
\begin{equation}
\sum_{n=1}^{\infty }C_{n}(x_{t},x_{n})=\frac{x_{t}(7-x_{t})}{16(x_{t}-1)}-%
\frac{\pi M_{w}Rx_{t}}{16(x_{t}-1)^{2}}[3(1+x_{t})J(R,-\frac{1}{2}%
)+(x_{t}-7)J(R,\frac{1}{2})].  \label{wilson15}
\end{equation}
$\bullet C_{10}$

$C_{10}$ is $\mu $ independent and is given by
\begin{equation}
C_{10}=-\frac{Y(x_{t},\frac{1}{R})}{\sin ^{2}\theta _{w}}.
\label{wilson16}
\end{equation}
The renormalization scale is fixed at $\mu =\mu _{b}\simeq 5$ GeV.

\subsection{Parameterizations of hadronic matrix element}

With the free quark decay amplitude available, we can proceed to
calculate the decay amplitudes for $\Lambda_b\to \Lambda \gamma$ and
$\Lambda_b \to \Lambda l^+l^-$ at hadron level, which can be
obtained by sandwiching the free quark amplitudes between the
initial and final baryon states in the spirit of factorization
assumption. Consequently, the following four hadronic matrix
elements
\begin{eqnarray}
\langle \Lambda(P)|\bar{s}\gamma_{\mu} b|\Lambda_{b}(P+q)\rangle
&,& \,\,\,  \langle \Lambda(P)|\bar{s}\gamma_{\mu}\gamma_5 b|\Lambda_{b}(P+q)\rangle ,
\nonumber \\
\langle \Lambda(P)|\bar{s}\sigma_{\mu \nu} b|\Lambda_{b}(P+q)\rangle
&,& \,\,\, \langle \Lambda(P)|\bar{s}\sigma_{\mu \nu} \gamma_5
b|\Lambda_{b}(P+q)\rangle,
\end{eqnarray}
need to be computed as can be observed from Eq. (\ref{effective
haniltonian 1}). Generally, the above four matrix elements can be
parameterized in terms of a series of form factors as \cite{c.q.
geng 4, Aliev 1,Aliev 2,Aliev 3,Aliev 4}
\begin{eqnarray}
\langle \Lambda(P)|\bar{s}\gamma_{\mu} b|\Lambda_{b}(P+q)\rangle
&=&\overline{\Lambda}(P)(g_1 \gamma_{\mu}+g_2 i \sigma_{\mu \nu}
q^{\nu}+g_3 q_{\mu})\Lambda_b(P+q), \,\, \label{vector
matrix element}\\
\langle \Lambda(P)|\bar{s}\gamma_{\mu}\gamma_5
b|\Lambda_{b}(P+q)\rangle &=&\overline{\Lambda}(P)(G_1
\gamma_{\mu}+G_2 i\sigma_{\mu \nu} q^{\nu}+G_3
q_{\mu})\gamma_{5}\Lambda_b(P+q), \,\, \label{axial-vector
matrix element}\\
\langle\Lambda(P)|\bar{s}i \sigma_{\mu \nu} q^{\nu}
b|\Lambda_b(P+q)\rangle &=&\overline{\Lambda}(P)(f_1
\gamma_{\mu}+f_2 i \sigma_{\mu \nu} q^{\nu}+f_3
q_{\mu})\Lambda_b(P+q),
\,\,\label{magnetic matrix element 1}\\
\langle\Lambda(P)|\bar{s}i \sigma_{\mu \nu}\gamma_{5} q^{\nu}
b|\Lambda_b(P+q)\rangle &=&\overline{\Lambda}(P)(F_1
\gamma_{\mu}+F_2 i \sigma_{\mu \nu} q^{\nu}+F_3
q_{\mu})\gamma_{5}\Lambda_b(P+q), \label{magnetic matrix element 2}
\end{eqnarray}
where all the form factors $g_i$, $G_i$, $f_i$ and $F_i$ are
functions of the square of momentum transfer $q^2$.

It should be emphasized that the form factors $f_3$ and $F_3$ do not
contribute to the decay amplitude of $\Lambda_b \to \Lambda + l^{+}
l^{-}$ due to the conservation of vector current, namely  $q^{\mu}
\bar{l} \gamma_{\mu} l = 0$.  Concentrating on the radiative decay
of $\Lambda_b\to \Lambda \gamma$, we then observe that the matrix
element of magnetic penguin operators can be simplified as
\begin{eqnarray}
\label{parameterization of tensor current 1}
\langle\Lambda(P)|\bar{s}i \sigma_{\mu \nu} q^{\nu}
b|\Lambda_b(P+q)\rangle &=& f_2(0) \overline{\Lambda}(P) i
\sigma_{\mu \nu} q^{\nu}\Lambda_b(P+q),
\,\,\\
\langle\Lambda(P)|\bar{s}i \sigma_{\mu \nu}\gamma_{5} q^{\nu}
b|\Lambda_b(P+q)\rangle &=&F_2(0) \overline{\Lambda}(P) i
\sigma_{\mu \nu}\gamma_5  q^{\nu}\Lambda_b(P+q).
\label{parameterization of tensor current 2}
\end{eqnarray}

For the completeness, we also present the parameterizations of
matrix elements involving the scalar $\bar{s} b$ and pseudo-scalar
$\bar{s} \gamma_5 b$ currents, which can be obtained from the Eqs.
(\ref{vector matrix element}) and (\ref{axial-vector matrix
element}) by contracting both sides to the four-momentum $q^{\mu}$
\begin{eqnarray}
\langle\Lambda(P)|\bar{s} b|\Lambda_b(P+q)\rangle &=& {1 \over
m_b+m_s} \overline{\Lambda}(P) [g_1(m_{\Lambda_b}-m_{\Lambda})+g_3
q^2]\Lambda_b(P+q), \,\, \label{scalar
matrix element}\\
\langle\Lambda(P)|\bar{s}\gamma_{5} b|\Lambda_b(P+q)\rangle &=& {1
\over m_b-m_s} \overline{\Lambda}(P)
[G_1(m_{\Lambda_b}+m_{\Lambda})-G_3 q^2]\gamma_5\Lambda_b(P+q).
\label{pseudo-scalar matrix element}
\end{eqnarray}

\section{Branching ratio, forward-backward asymmetry and polarization asymmetry}

Now, we are going to analyze the sensitivity of  the branching
ratio, forward-backward asymmetry and polarization asymmetry of
$\Lambda$ baryon on the radius of extra dimension $R$. To this
purpose, we firstly list the input parameters used in this paper in
Table \ref{Input parameters}.

\begin{table}[tbh]
\caption{{}Values of input parameters used in our numerical
analysis} \label{Input parameters}
\begin{tabular}{cc}
\hline\hline $G_{F}=1.166\times 10^{-2}$ GeV$^{-2}$ & $\left|
V_{ts}\right|
=41.61_{-0.80}^{+0.10}\times 10^{-3}$ \\
$\left| {V_{tb}}\right| =0.9991$ & $m_{b}=\left( 4.68\pm 0.03\right)
$ GeV
\\
{$m_{c}\left( m_{c}\right) =1.275_{-0.015}^{+0.015}$ GeV} & $m_{s}\left( 1%
\text{ GeV}\right) =\left( 142\pm 28\right) $ MeV \\ \hline
$m_{\Lambda _{b}}=5.62$ GeV & $m_{\Lambda }=1.12$ GeV \\ \hline
$m_{\mu}=0.106 {\rm GeV}$ &$m_{\tau}=1.777 {\rm GeV}$ \\
\hline\hline
\end{tabular}
\end{table}

In addition, we also collect here the form factors calculated in the
Ref. \cite{Wang:2008sm}, where the effects of higher twist
distribution amplitudes of $\Lambda$ baryon are included in the sum
rules of transition form factors. Specifically, the dependence of
form factors on the transfer momentum are parameterized as
\cite{Wang:2008sm}
\begin{equation}
\xi _{i}(q^{2})={\frac{\xi _{i}(0)}{1-a_{1}q^{2}/m_{\Lambda
_{b}}^{2}+a_{2}q^{4}/m_{\Lambda _{b}}^{4}}}, \label{pole model of
form factors}
\end{equation}%
where $\xi _{i}$ denotes the form factors $f_{2}$ and $g_{2}$. The
numbers
of parameters $\xi_i(0), \,\, a_1, \,\, a_2$ have been collected in Table %
\ref{di-fit}.
\begin{table}[tbh]
\caption{{}Numerical results for the form factors $f_{2}(0)$,
$g_{2}(0)$ and parameters $a_{1}$ and $a_{2}$ involved in the
double-pole fit of Eq. (\ref {pole model of form factors}) for both
twist-3 and twist-6 sum rules with $M_{B}^{2} \in \lbrack
3.0,6.0]~\mbox{GeV}^{2}$, $s_{0}=39\pm 1~\mbox{GeV}^{2}$.}
\begin{tabular}{cccccc}
\hline \hline
parameter  & {COZ}&  {FZOZ} &QCDSR & twist-3  & up to twist-6\\
\hline $f_2(0)$ & $0.74^{+0.06}_{-0.06}$ & $0.87^{+0.07}_{-0.07}$ &
$0.45$ &$0.14^{+0.02}_{-0.01}$
&$0.15^{+0.02}_{-0.02}$\\
{$a_1$} & $2.01^{+0.17}_{-0.10}$  & $2.08^{+0.15}_{-0.09}$ &0.57
&$2.91^{+0.10}_{-0.07}$
&$2.94^{+0.11}_{-0.06}$\\
{$a_2$}& $1.32^{+0.14}_{-0.08}$ &$1.41^{+0.11}_{-0.08}$&
$-0.18$&$2.26^{+0.13}_{-0.08}$
&$2.31^{+0.14}_{-0.10}$ \\
\hline
 $g_2 (0) (10^{-2}\rm{GeV^{-1}})$  & $-2.4^{+0.3}_{-0.2}
 $ & $-2.8^{+0.4}_{-0.2}  $ & $-1.4$ &$-0.47^{+0.06}_{-0.06}
 $ &$1.3^{+0.2}_{-0.4} $
\\
 {$a_1$}& $2.76^{+0.16}_{-0.13}$ & $2.80^{+0.16}_{-0.11}$ &2.16 &$3.40^{+0.06}_{-0.05}$
 &$2.91^{+0.12}_{-0.09}$\\
{$a_2$} &$2.05^{+0.23}_{-0.13}$ &$2.12^{+0.21}_{-0.13}$ &1.46
&$2.98^{+0.09}_{-0.08}$
&$2.24^{+0.17}_{-0.13}$ \\
\hline \hline \label{di-fit}
\end{tabular}
\end{table}
To the leading order in $\alpha_s$ and leading contributions in the
infinite momentum kinematics, the other form factors can be related
to these two as
\begin{eqnarray}
F_1(q^2)&=&f_1(q^2)=q^2 g_2(q^2)=q^2 G_2(q^2),  \notag \\
F_2(q^2)&=&f_2(q^2)=g_1(q^2)=G_1(q^2),
\end{eqnarray}
where the form factors $F_3(q^2)$ and $G_3(q^2)$ are dropped out
here due to their tiny contributions.

\subsection{Decay width of $\Lambda_b \to \Lambda + \gamma$}

\begin{figure}[tb]
\begin{center}
\begin{tabular}{ccc}
\includegraphics[scale=0.6]{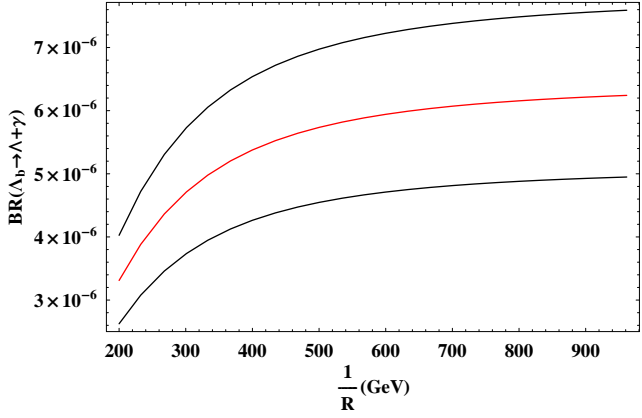}
\includegraphics[scale=0.6]{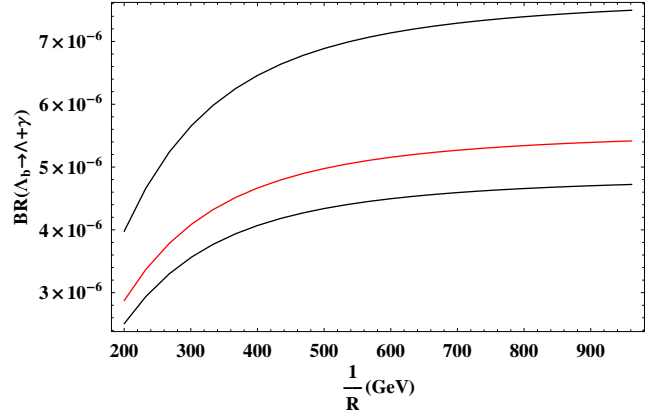}
\\
\includegraphics[scale=0.6]{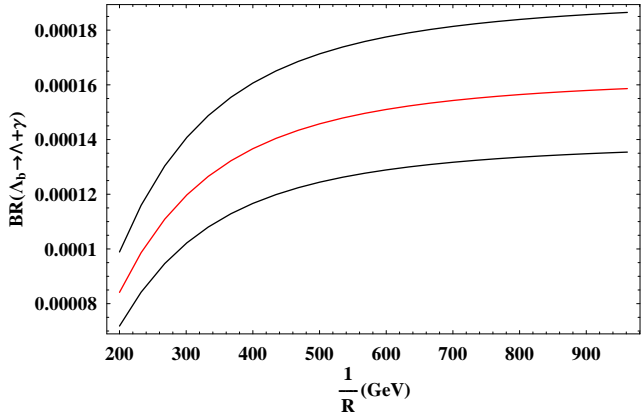}
\includegraphics[scale=0.6]{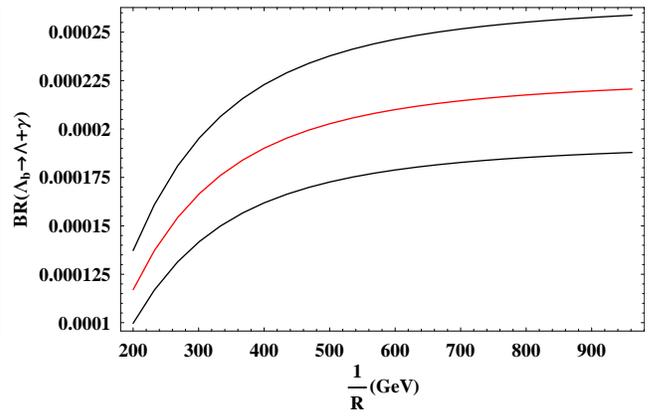}
\put (-350,290){(a)} \put (-100,290){(b)} \put (-350,0){(c)}
\put(-100,0){(d)}
\end{tabular}
\caption{(Color online). Decay rate of $\Lambda_b \to \Lambda +
\gamma$ as a function of compactification parameter $1/R$. The
center line describes the case using the central values of form
factors; while the other two lines correspond to the results of
branching ratio when the errors in the form factors are added and
subtracted from their central values. The fig. (a) and (b) denote
the $BR(\Lambda_b \to \Lambda + \gamma)$ using the transition form
factors up to twist-6 and twist-3 respectively; while fig. (c) and
(d) are obtained based on the form factors calculated in the COZ and
FZOZ model respectively.}\label{BR of radiative decay}
\end{center}
\end{figure}

Making use of Eqs. (\ref{parameterization of tensor current 1}) and
(\ref{parameterization of tensor current 2}), the decay width of
$\Lambda_b \to \Lambda \gamma$ can be written as
\begin{eqnarray}
\Gamma(\Lambda_b \to \Lambda \gamma) = {\alpha_{em} G_F^2 \over 32
m_{\Lambda_b}^3\pi^4}|V_{tb}|^2|V_{ts}|^2 |C_7|^2 (1-x^2)^3
(m_b^2+m_s^2)[f_2(0)]^2 . \label{decay width of radiative decay}
\end{eqnarray}
The numerical evaluation of $BR(\Lambda_b \to \Lambda \gamma)$ in
the SM indicates that it can be as large as $(0.63-0.73) \times
10^{-5}$, hence  this process is  within the reach of LHC
experiments. To explore the effects of KK states in this process, we
present the dependence of decay rate on the compactification
parameter $1/R$ in Fig. \ref{BR of radiative decay}.  As can be
observed from this figure, the branching fraction is suppressed for
low values of $1/R$, which is around 25\% smaller for $1/R=300
{\rm{GeV}}$ than that in the SM. Such kind of suppression from KK
modes is reflected in the Wilson coefficient $C_7^{(0)eff}(\mu_b)$,
whose value at $m_b$ scale decreases from $-0.300$ in the SM to
$-0.245$ in the ACD model \cite{Buras:2003mk}. As a matter of fact,
the suppression of branching fraction for $b \to s$ transition has
already been found in the inclusive decay process $B \to X_s \gamma$
\cite{Buras:2003mk,Agashe:2001ra}.

For a comparison, we also display the result of $BR(\Lambda_b\to
\Lambda \gamma)$ as a function of parameter $1/R$ in Fig. \ref{BR of
radiative decay}, utilizing the transition form factors calculated
in the COZ and FZOZ modes for the distribution amplitudes of
$\Lambda$ baryon. Up to now, only the upper bound $1.3 \times
10^{-3}$ for branching ratio of $\Lambda_b\to \Lambda \gamma$ decay
is available in experiment, so the forthcoming experimental data can
not only be used to constrain the additional parameter $R$ with
respect to the SM but also are helpful to discriminate existing
models of distribution amplitudes for $\Lambda$ baryon.

\subsection{Decay width and dilepton distributions of $\Lambda_b \to \Lambda + l^{+} l^{-}$}

The differential decay width of $\Lambda_b \to \Lambda l^+ l^-$ in
the rest frame of $\Lambda_b$ baryon can be written as \cite{PDG},
\begin{equation}
{d\Gamma({\Lambda_b \to \Lambda l^+ l^-}) \over d q^2} ={1 \over (2
\pi)^3} {1 \over 32 m_{\Lambda_b}^3} \int_{u_{min}}^{u_{max}}
|{\widetilde{M}}_{\Lambda_b \to \Lambda l^+ l^-}|^2 du,
\label{differential decay width}
\end{equation}
where $u=(p_{\Lambda}+p_{l^{-}})^2$ and $q^2=(p_{l^+}+p_{l^-})^2$;
$p_{\Lambda}$, $p_{l^{+}}$ and $p_{l^{-}}$ are the four-momenta
vectors of $\Lambda$, $l^{+}$ and $l^{-}$ respectively.
${\widetilde{M}}_{\Lambda_b \to \Lambda l^+ l^-}$ is the  decay
amplitude after integrating over the angle between the $l^{-}$ and
$\Lambda$ baryon. The upper and lower limits of $u$ are given by
\begin{eqnarray}
u_{max}&=&(E^{\ast}_{\Lambda}+E^{\ast}_{l})^2-(\sqrt{E_{\Lambda}^{\ast
2}-m_{\Lambda}^2}-\sqrt{E_l^{\ast 2}-m_l^2})^2, \nonumber\\
u_{min}&=&(E^{\ast}_{\Lambda}+E^{\ast}_{l})^2-(\sqrt{E_{\Lambda}^{\ast
2}-m_{\Lambda}^2} +\sqrt{E_l^{\ast 2}-m_l^2})^2;
\end{eqnarray}
where $E^{\ast}_{\Lambda}$ and $E^{\ast}_{l}$ are the energies of
$\Lambda$ and $l^{-}$ in the rest frame of lepton pair
\begin{equation}
E^{\ast}_{\Lambda}= {m_{\Lambda_b}^2 -m_{\Lambda}^2 -q^2 \over 2
\sqrt{q^2}}, \hspace {1 cm} E^{\ast}_{l}={q^2\over 2\sqrt{q^2}}.
\end{equation}

The total decay rates of $\Lambda_b \to \Lambda l^+ l^-$ ($l=\mu, \,
\tau$) in the ACD model have been plotted in the Fig. \ref{BR of
semileptonic decay}, from which we can observe that the KK states
can result in 10\% enhancement with fixed $1/R=300 {\rm GeV}$
compared with that in the SM. This can be easily understood since
the Wilson coefficient $C_9$ is essentially the same as that in the
SM and $C_{10}$ is significantly enhanced, which can overwhelm the
suppression from $C_7$. The enhancement effect due to KK modes in
the $b \to s l^{+} l^{-}$ transition is already found in the
inclusive decay $B \to X_s \mu^{+} \mu^{-}$ \cite{Buras:2003mk}.
Furthermore, the dilepton distributions  of $\Lambda_b \to \Lambda +
l^{+} l^{-}$ are also displayed in Fig. \ref{Dilepton distributions
of semileptonic decay}, where the predictions in the SM are also
included for completeness. As can be seen, the  invariant mass
distribution amplitude is not sensitive to the effect of extra
dimension  for both the muon and tauon  cases.

\begin{figure}[tb]
\begin{center}
\begin{tabular}{ccc}
\includegraphics[scale=0.6]{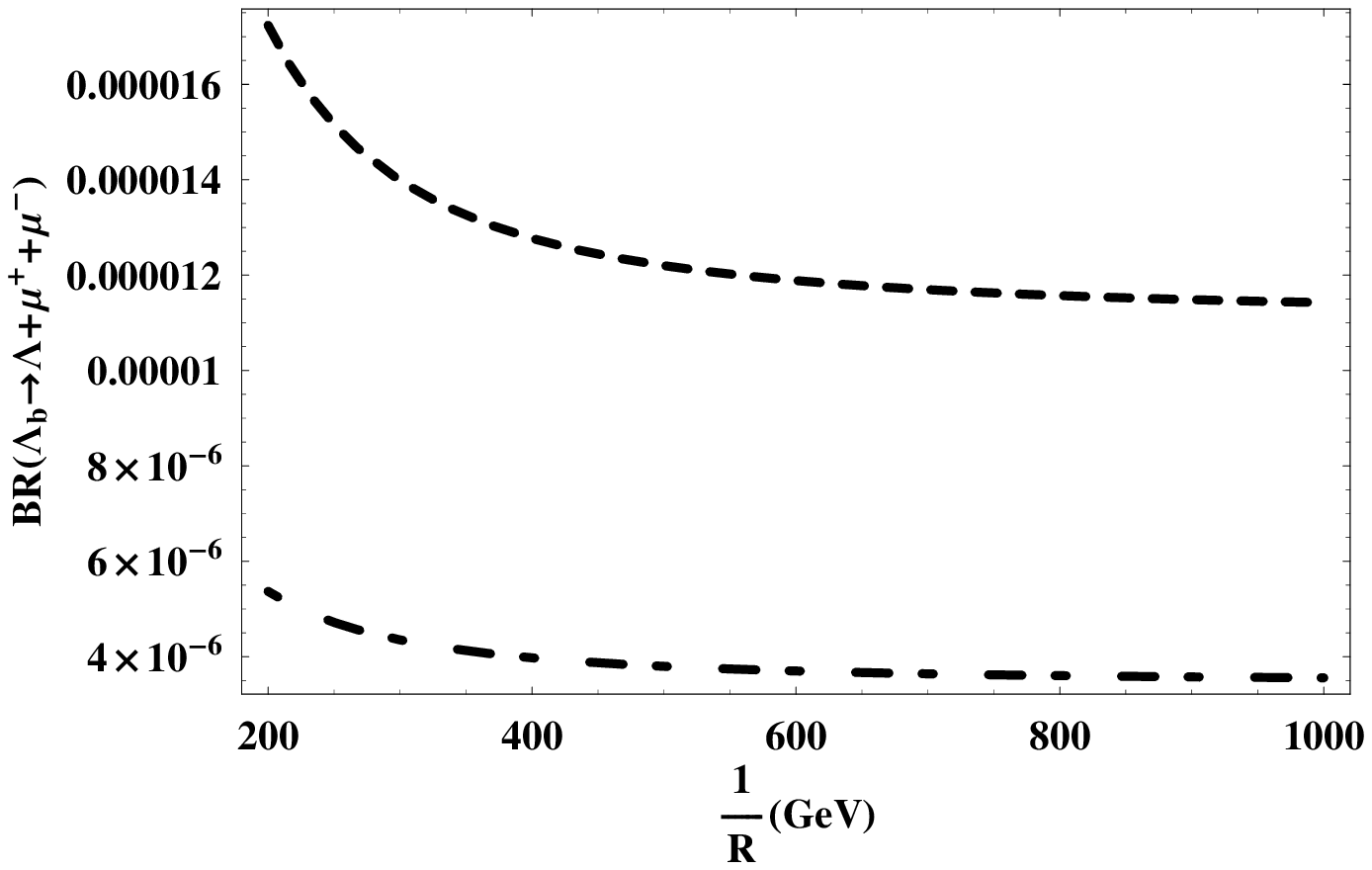}
\includegraphics[scale=0.6]{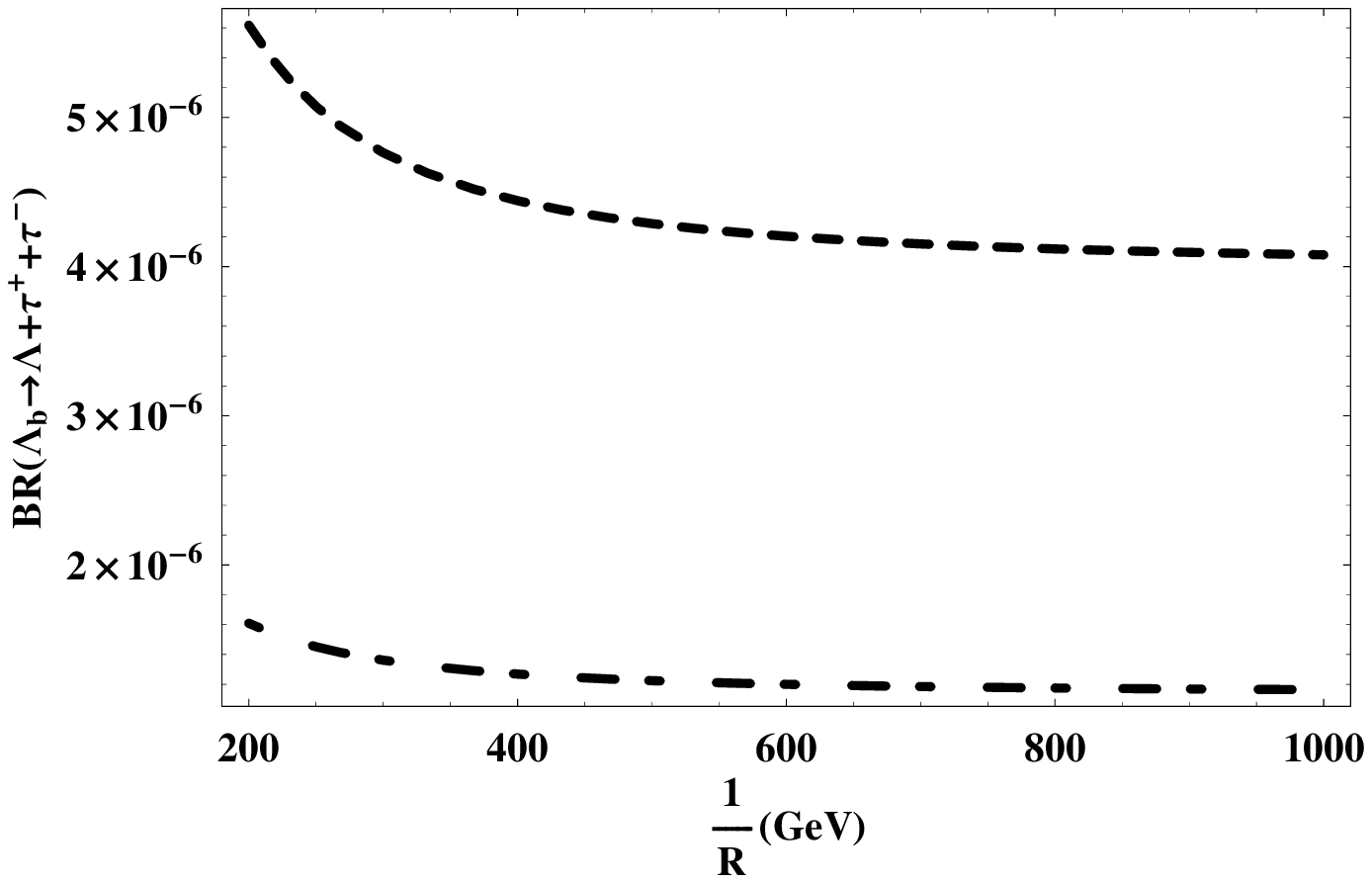}
\end{tabular}
\caption{Decay rate of $\Lambda_b \to \Lambda + l^{+} l^{-}$
($l=\mu, \, \tau$) as a function of compactification parameter $1/R$
within the range [200,1000] GeV. The dashed (dot-dashed) lines
correspond to the results of branching fractions when the errors in
the form factors are added to (subtracted from) their central
values. }\label{BR of semileptonic decay}
\end{center}
\end{figure}

\begin{figure}[tb]
\begin{center}
\begin{tabular}{ccc}
\includegraphics[scale=0.6]{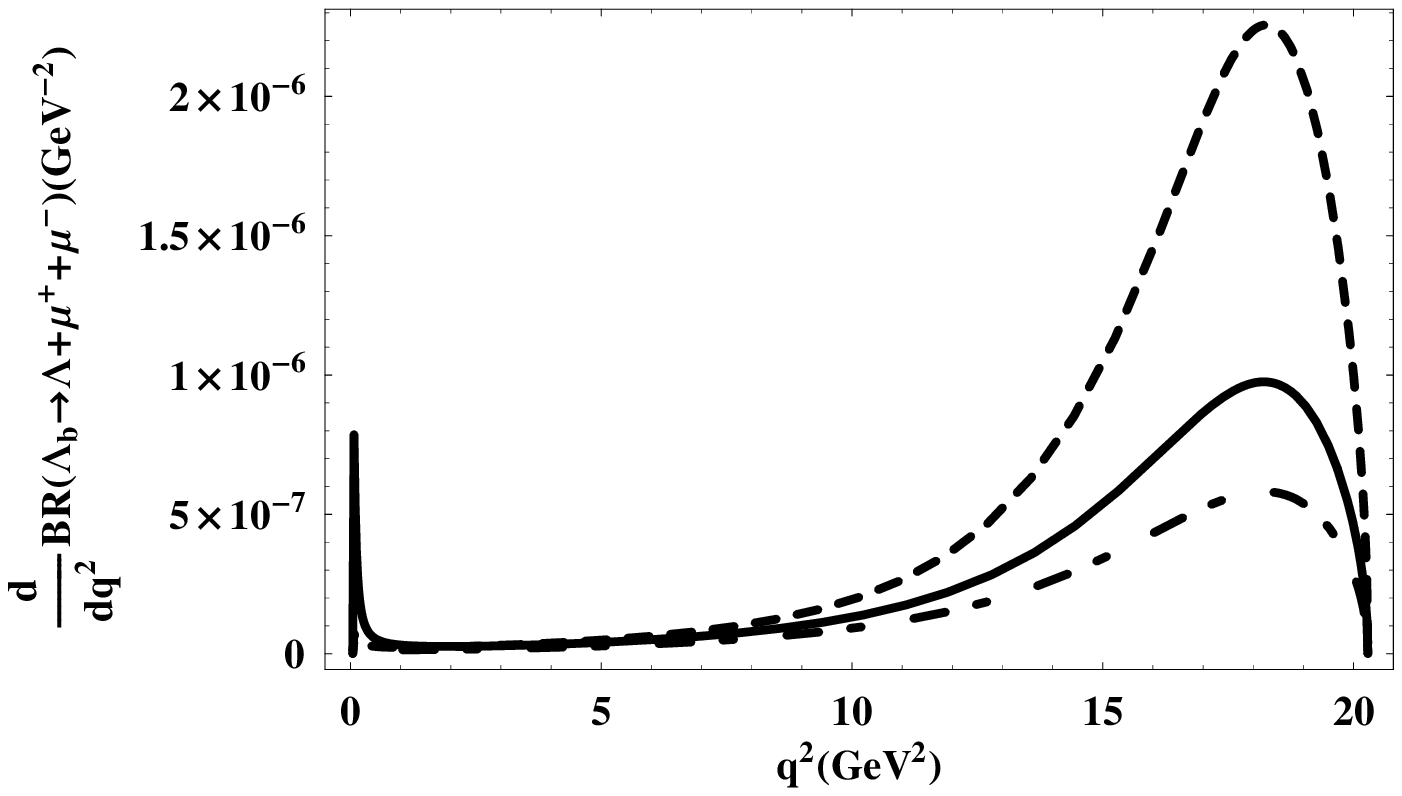}
\includegraphics[scale=0.6]{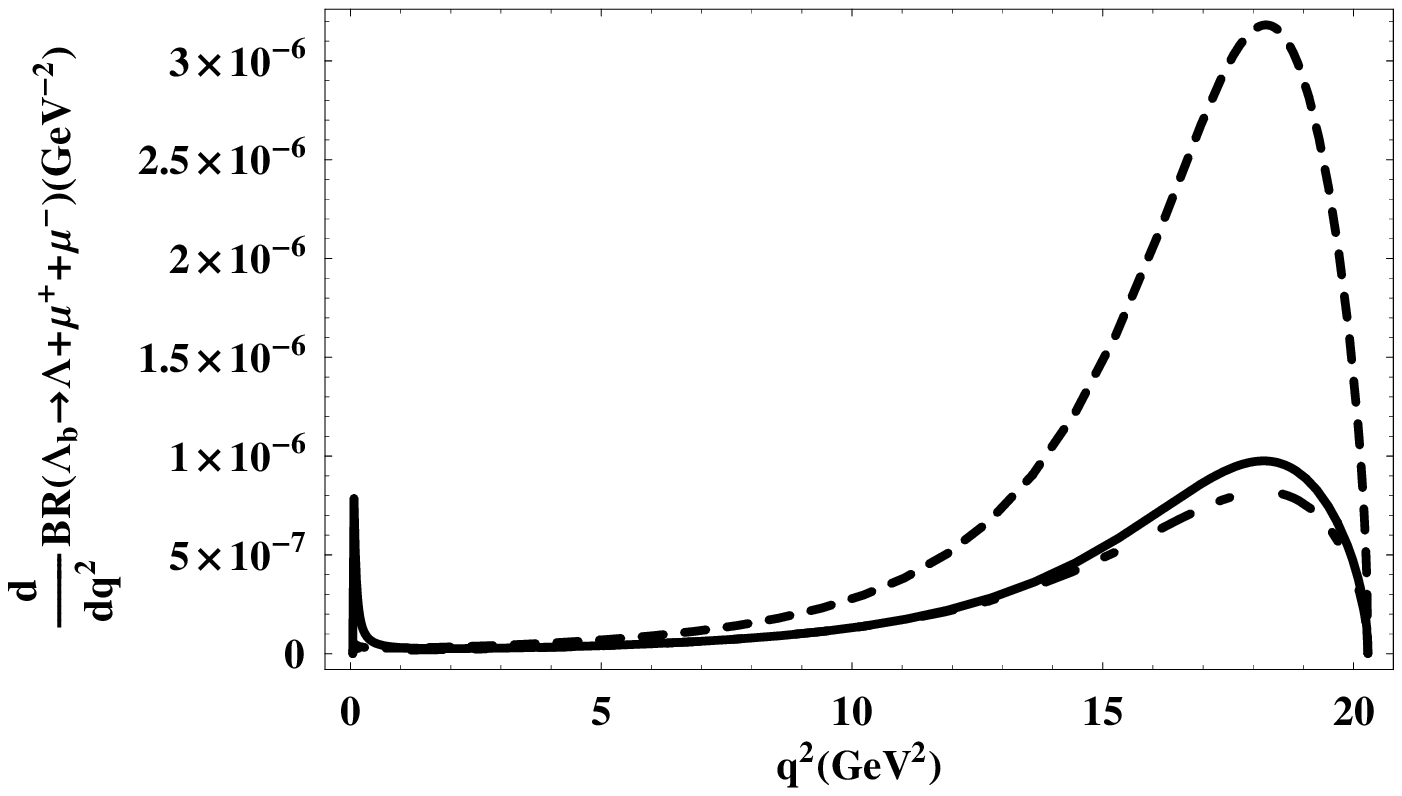}
\\
\includegraphics[scale=0.6]{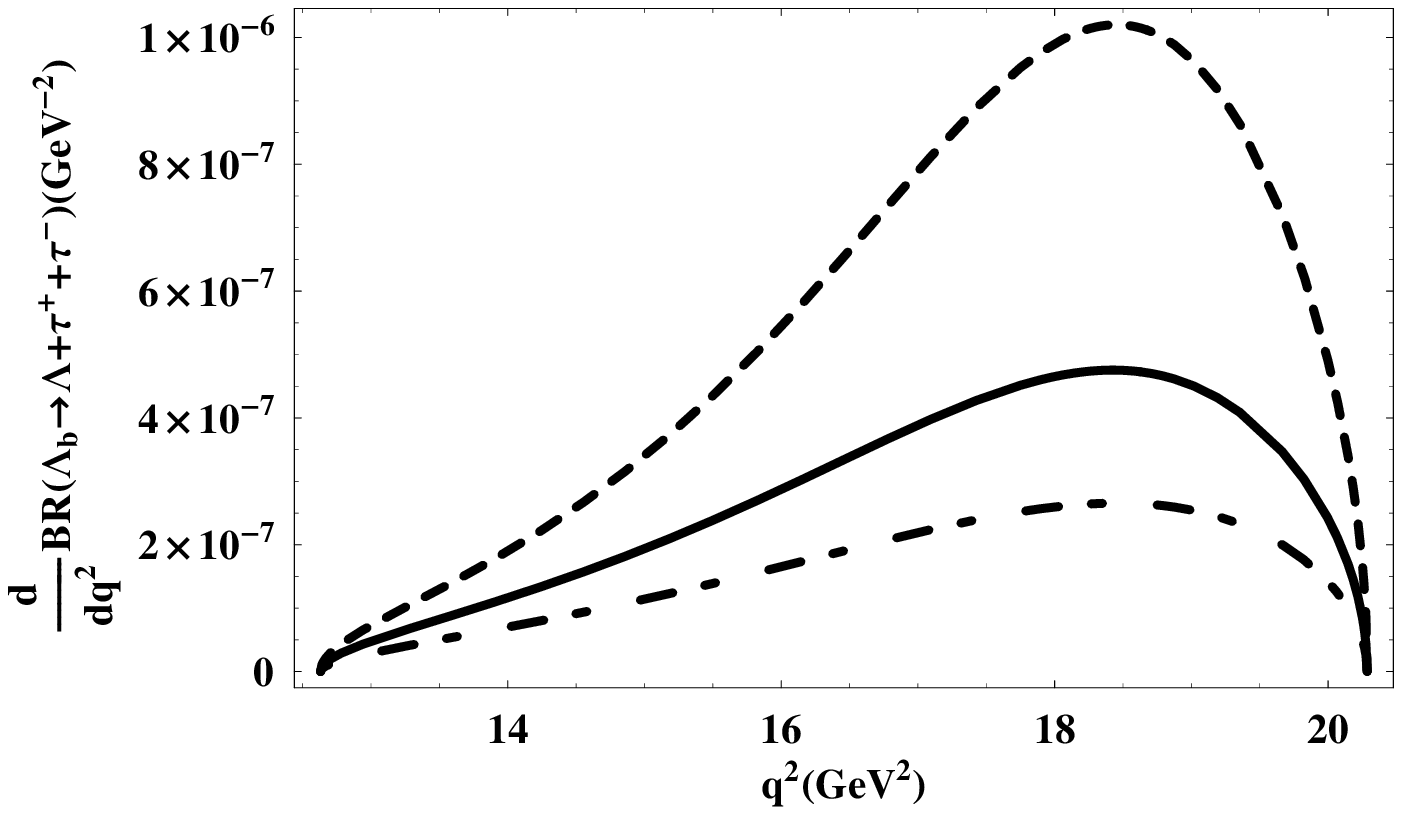}
\includegraphics[scale=0.6]{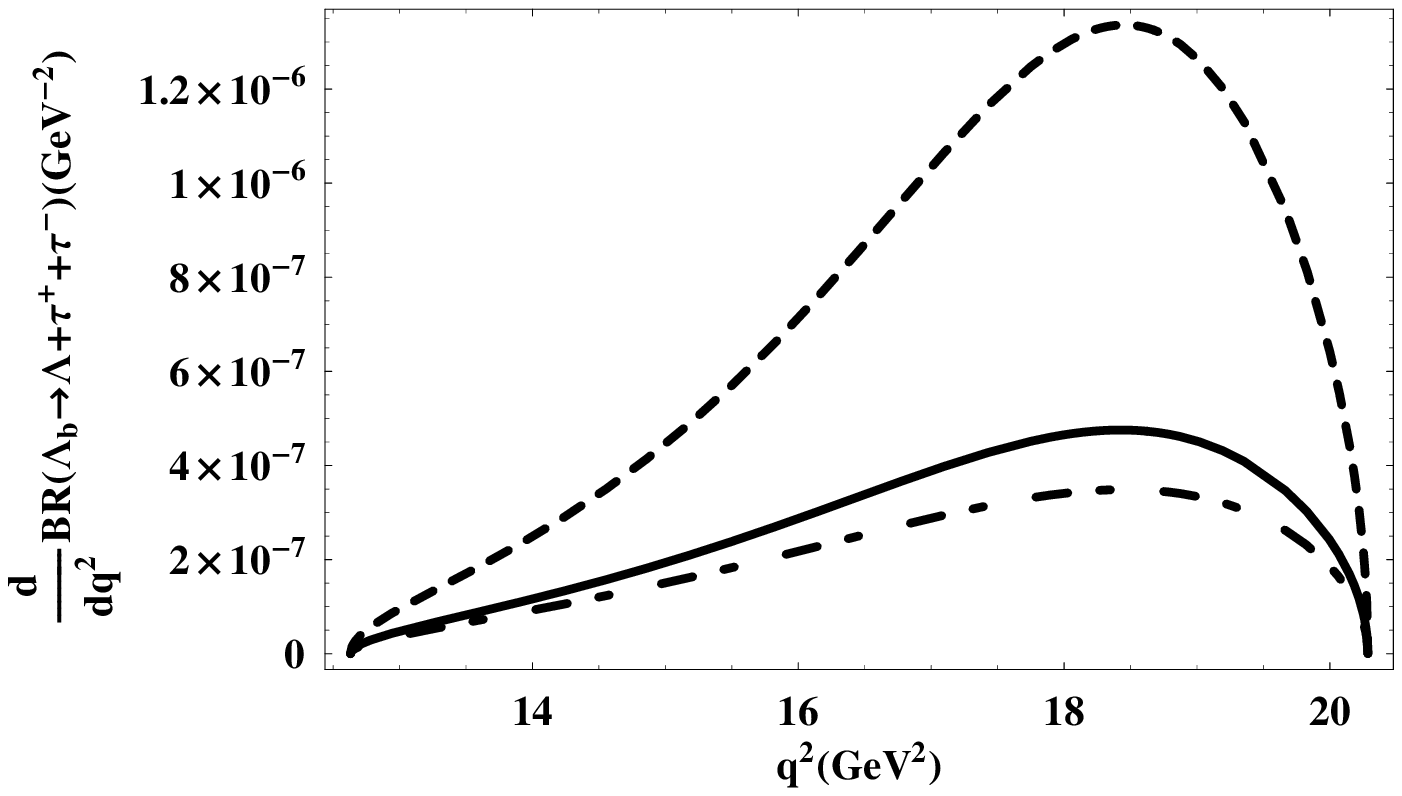}
\put (-350,290){(a)} \put (-100,290){(b)} \put (-350,0){(c)}
\put(-100,0){(d)}
\end{tabular}
\caption{Dilepton distributions  of $\Lambda_b \to \Lambda +  l^{+}
l^{-}$ as a function of $q^{2}$ at two different values of $1/R$.
The fig. (a) and (b) are for the muon case with $1/R=500 \rm{GeV}$
and $1/R=200 \rm{GeV}$ respectively;  while fig. (c) and (d) are for
the tauon case with $1/R=500 \rm{GeV}$  and $1/R=200 \rm{GeV}$
respectively. The solid lines correspond to the results obtained in
the SM with the central value of form factors, the dashed
(dot-dashed) lines describe the results in the ACD model when the
errors in the form factors are added to (subtracted from) their
central values.}\label{Dilepton distributions of semileptonic decay}
\end{center}
\end{figure}

\subsection{Forward-backward asymmetry of $\Lambda_b \to \Lambda + l^{+} l^{-}$}

Following Refs. \cite{c.q. geng 4, b to s in theory 9}, the
differential and normalized forward-backward asymmetries for the
semi-leptonic decay $\Lambda_b \to \Lambda l^{+} l^{-}$ can be
defined as
\begin{eqnarray}
{d A_{FB}(q^2) \over d q^2}=\int_0^1 dz  {d^2 \Gamma (q^2, z) \over
dq^2 dz} - \int_{-1}^0 dz  {d^2 \Gamma (q^2, z) \over dq^2 dz}.
\end{eqnarray}
and
\begin{eqnarray}
A_{FB}(q^2)={ \int_0^1 dz  {d^2 \Gamma (q^2, z) \over dq^2 dz} -
\int_{-1}^0 dz  {d^2 \Gamma (q^2, z) \over dq^2 dz}\over \int_0^1 dz
{d^2 \Gamma (q^2, z) \over dq^2 dz} + \int_{-1}^0 dz  {d^2 \Gamma
(q^2, z) \over dq^2 dz} }.
\end{eqnarray}
Making use of the decay amplitude in Eq. (\ref{b to s l l}), the
differential forward-backward asymmetry for decays of $\Lambda_b \to
\Lambda + l^{+} l^{-}$ can be calculated as

\begin{eqnarray}
{d A_{FB}(q^2) \over d q^2}={ G_F^2 \alpha_{em}^2 |V_{tb}
V_{ts}^{\ast}|^2 \over 256 m_{\Lambda_b}^3 \pi^5}
\lambda(m_{\Lambda_b}^2,m_{\Lambda}^2,q^2) (1-{4 m_l^2 \over q^2})
R_{FB}(q^2),
\end{eqnarray}
with
\begin{eqnarray}
R_{FB}(q^2)&=&2[(m_s m_{\Lambda}+m_b m_{\Lambda_b}) f_2^2- m_s
(m_{\Lambda}^2-m_{\Lambda_b}^2+q^2) f_2 g_2  + (m_s m_{\Lambda}-m_b
m_{\Lambda_b}) q^2 g_2^2] {\rm {Re}}(C_7^{eff} C_{10}^{\ast}) \nonumber \\
&&+[(f_2 -g_2 m_{\Lambda})^2-g_2^2 m_{\Lambda_b}^2]q^2 {\rm
{Re}}(C_9^{eff} C_{10}^{\ast}), \label{FBA expressions}
\end{eqnarray}
where we have retained masses for both the lepton and strange quark.

In fig. \ref{forward-backward asymmetry of semileptonic decay}, we
show the dependence of $A_{FB}$ on the momentum transfer $q^2$ at
two fixed values of $1/R=200 {\rm GeV}$ and $500 {\rm GeV}$ as well
as that in the  SM for both the muon and tauon cases. As can be seen
from the figure, the zero-position of forward-backward asymmetry for
$\Lambda_b \to \Lambda + \mu^{+} \mu^{-}$ is sensitive on the
compactification parameter $1/R$, which is consistent with that
observed in \cite{Aliev:2006xd}. For the case of $1/R=500 {\rm
GeV}$, the forward-backward asymmetry is quite close to that in the
SM for both two cases of the final states. For this reason,
experimental determination of the zero-point of $A_{FB}$ for
$\Lambda_b \to \Lambda + \mu^{+} \mu^{-}$ can provide valuable
information on the new physics effects. Similar to that in the SM,
there is no zero position of forward-backward asymmetry for the case
of $\Lambda_b \to \Lambda + \tau^{+} \tau^{-}$ in the ACD model
apart from the end point regions.

\begin{figure}[tb]
\begin{center}
\begin{tabular}{ccc}
\includegraphics[scale=0.6]{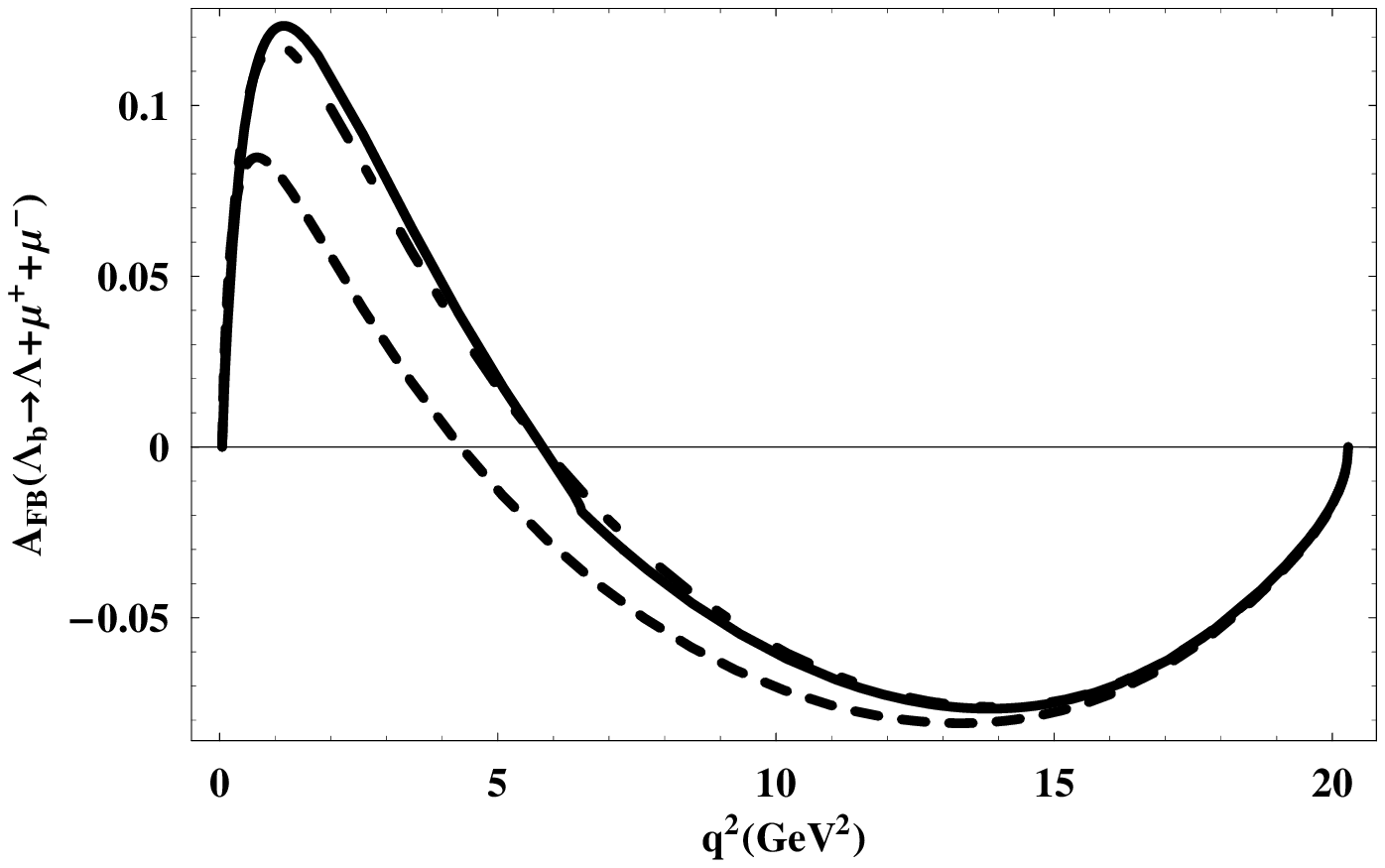}
\includegraphics[scale=0.6]{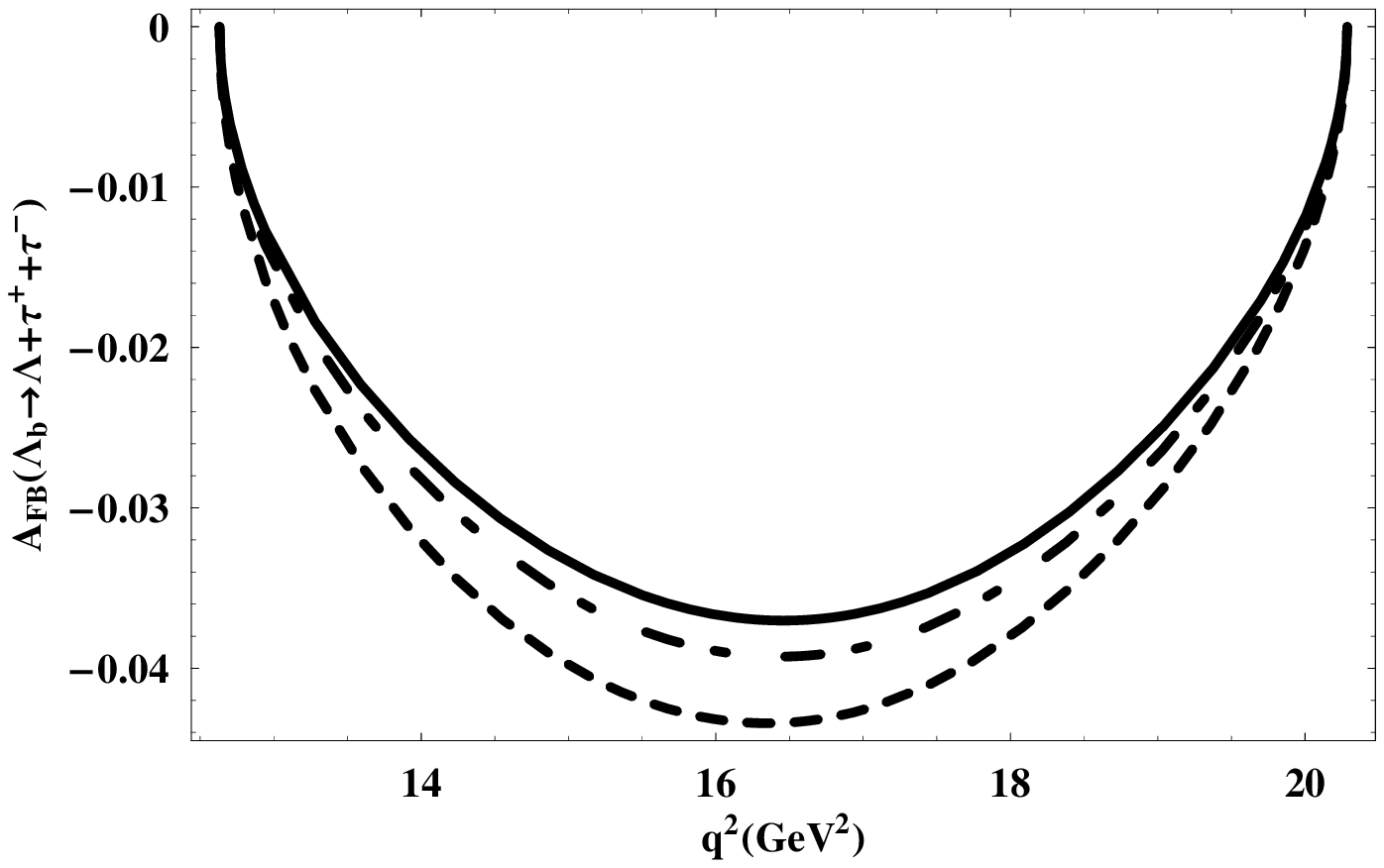}
\end{tabular}
\caption{Forward-backward asymmetry  of $\Lambda_b \to \Lambda +
l^{+} l^{-}$ ($l=\mu, \, \tau$) as a function of momentum transfer
$q^2$ at two fixed value of $1/R$ and in the SM. The solid line
represents  the case in the SM; while the dashed (dot-dashed) lines
correspond to the cases with the value of $1/R$ being 200 GeV and
500 GeV respectively. }\label{forward-backward asymmetry of
semileptonic decay}
\end{center}
\end{figure}

\subsection{$\Lambda$ baryon polarization asymmetry of $\Lambda_b \to \Lambda + l^{+} l^{-}$}

To study the $\Lambda $ spin polarization, one needs to express the
$\Lambda $ four spin vector in terms of a unit vector $\vec{\xi}$
along the $\Lambda $
spin in its rest frame as \cite{c.q. geng 2}%
\begin{equation}
s_{0}=\frac{\vec{p}_{\Lambda }\cdot \vec{\xi}}{m_{\Lambda }},\text{ }\vec{s}=%
\vec{\xi}+\frac{s_{0}}{E_{\Lambda }+m_{\Lambda }}\vec{p}_{\Lambda }.
\label{polarization-lambda}
\end{equation}
The unit vectors along the longitudinal, normal and transverse
components of the $\Lambda $ polarization are chosen to be
\begin{eqnarray}
\hat{e}_{L} &=&\frac{\vec{p}_{\Lambda }}{\left| \vec{p}_{\Lambda
}\right| },
\\
\hat{e}_{N} &=&\frac{\vec{p}_{\Lambda }\times (\vec{p}_{-}\times \vec{p}%
_{\Lambda })}{\left| \vec{p}_{\Lambda }\times (\vec{p}_{-}\times \vec{p}%
_{\Lambda })\right| }, \\
\hat{e}_{T} &=&\frac{\vec{p}_{-}\times \vec{p}_{\Lambda }}{\left| \vec{p}%
_{-}\times \vec{p}_{\Lambda }\right| },
\end{eqnarray}
where $\vec{p}_{-}$ and $\vec{p}_{\Lambda }$  are the three-momenta
of the lepton $l^{-}$ and $\Lambda $ baryon respectively in the
center mass frame of $l^{+}l^{-}$ system.

The polarization asymmetries for $%
\Lambda $ baryon in $\Lambda _{b}\rightarrow \Lambda l^{+}l^{-}$ can
be defined as
\begin{equation}
P_{i}^{(\mp )}(q^2)=\frac{\frac{d\Gamma }{d
q^2}(\vec{\xi}=\hat{e}_i)-\frac{d\Gamma
}{d q^2}(\vec{\xi}=-\hat{e}_i)}{\frac{d\Gamma }{d q^2}(\vec{\xi}=\hat{e}_i)+\frac{%
d\Gamma }{d q^2}(\vec{\xi}=-\hat{e}_i)}  \label{LNT for Lambda}
\end{equation}%
where $i=L,\;N,\;T$ and $\vec{\xi}$ is the spin direction along the
$\Lambda
$ baryon. The differential decay rate for polarized $\Lambda $ baryon in $%
\Lambda _{b}\rightarrow \Lambda l^{+}l^{-}$ decay along any spin direction $%
\vec{\xi} $ is related to the unpolarized decay rate
(\ref{differential
decay width}) through the following relation%
\begin{equation}
\frac{d\Gamma (\vec{\xi})}{d q^2}=\frac{1}{2}\left( \frac{d\Gamma
}{d q^2}\right)
[1+(P_{L}\vec{e}_{L}+P_{N}\vec{e}_{N}+P_{T}\vec{e}_{T})\cdot
\vec{\xi}]. \label{decay rate polarized Lambda}
\end{equation}

In the Fig. \ref{longitudinal polarization asymmetry of semileptonic
decay}, we display the longitudinal polarization asymmetry of
$\Lambda$ baryon for both the muon and tauon cases with two fixed
numbers of $1/R$ in the ACD model together with that in the SM, from
which one can see that the impact of extra dimension on this
asymmetry is rather weak. The normal polarization asymmetry of
$\Lambda$ baryon has been plotted in fig. \ref{normal polarization
asymmetry of semileptonic decay}, from which we can find that the
effects of KK states are more important at large momentum transfer
$q^2$ and might be distinguishable from that in the SM for the case
of $1/R=200 \rm{GeV}$. As for the transverse polarization asymmetry,
both the ACD model and the SM can give very tiny predictions,
which is almost impossible to detect in the future colliders. In
short, the measurement of polarization asymmetries of $\Lambda$
baryon in $\Lambda_b \to \Lambda + l^{+} l^{-}$ decays is not so
helpful to establish the UED models.  This is much very similar to
the case of single-lepton polarization as found in Ref.
\cite{Aliev:2006xd}.

\begin{figure}[tb]
\begin{center}
\begin{tabular}{ccc}
\includegraphics[scale=0.6]{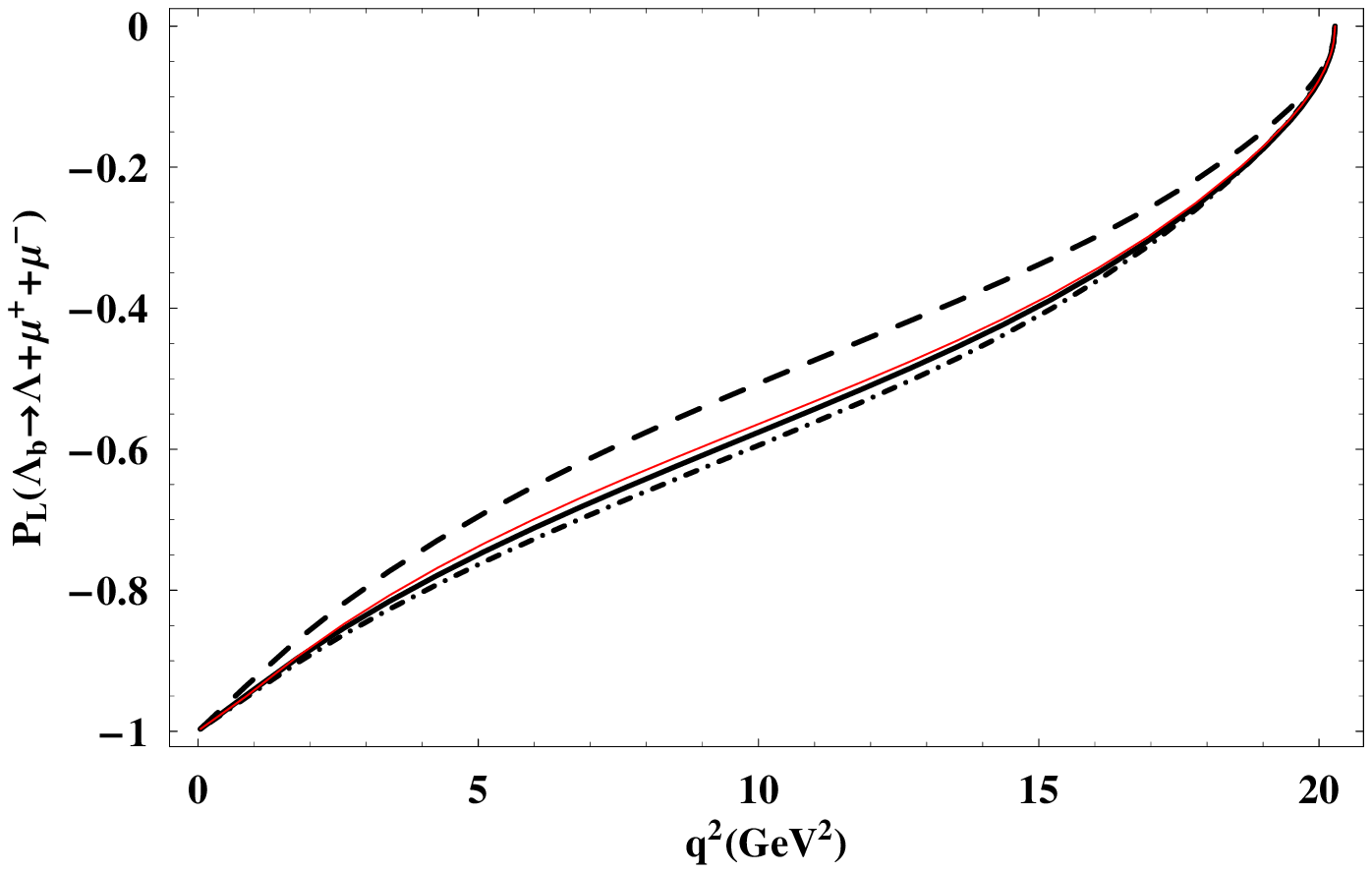}
\includegraphics[scale=0.6]{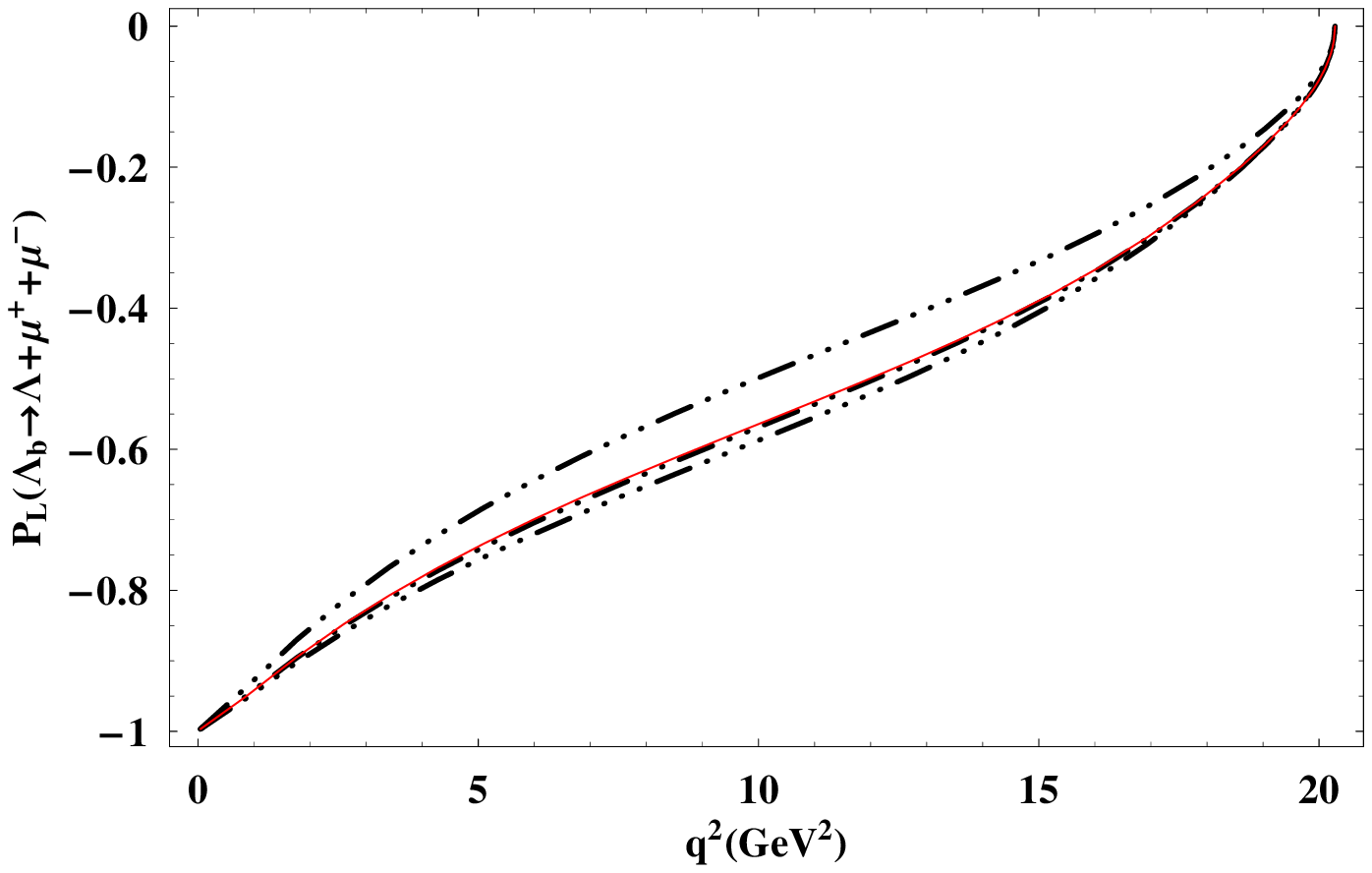}
\\
\includegraphics[scale=0.6]{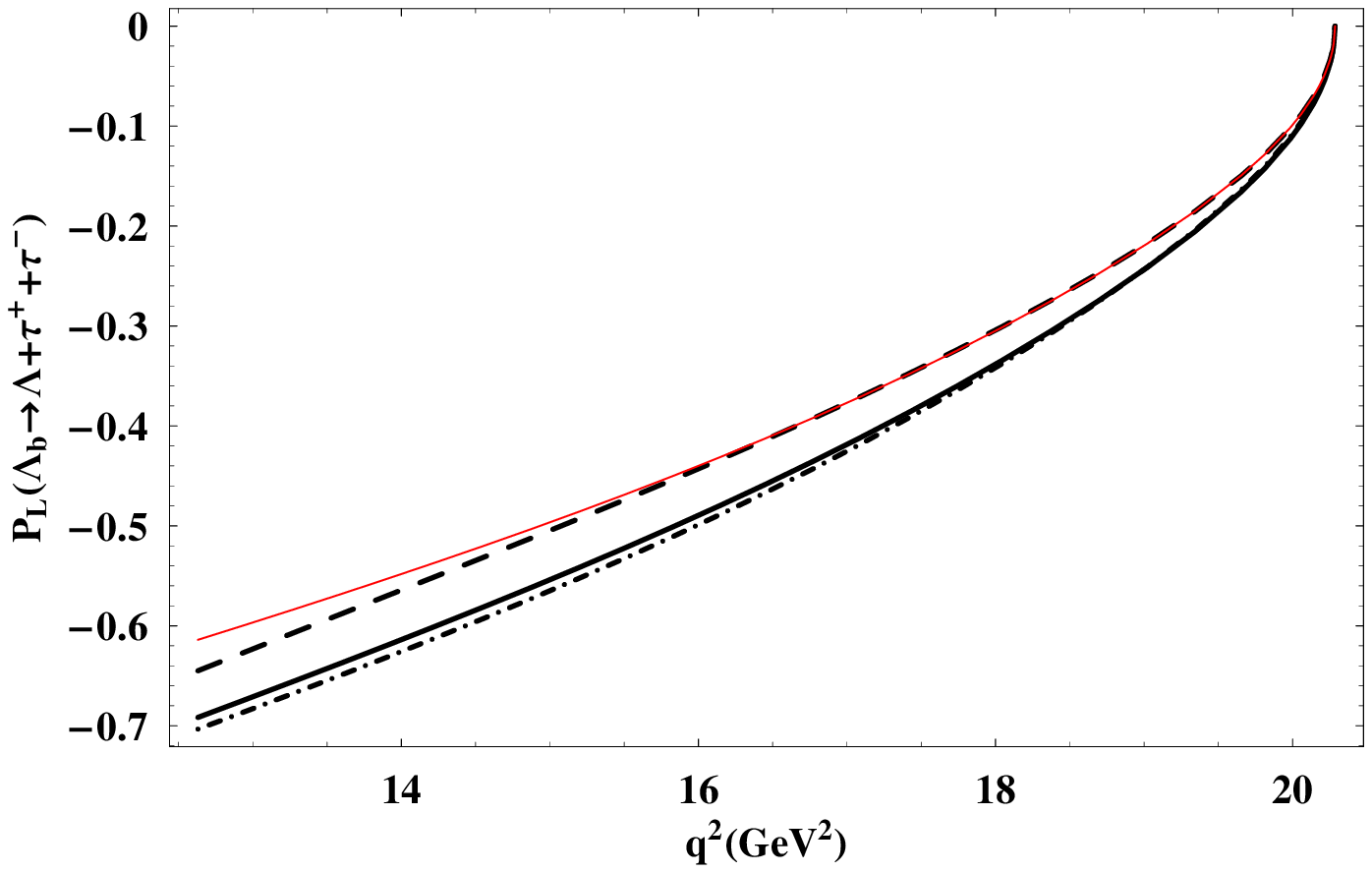}
\includegraphics[scale=0.6]{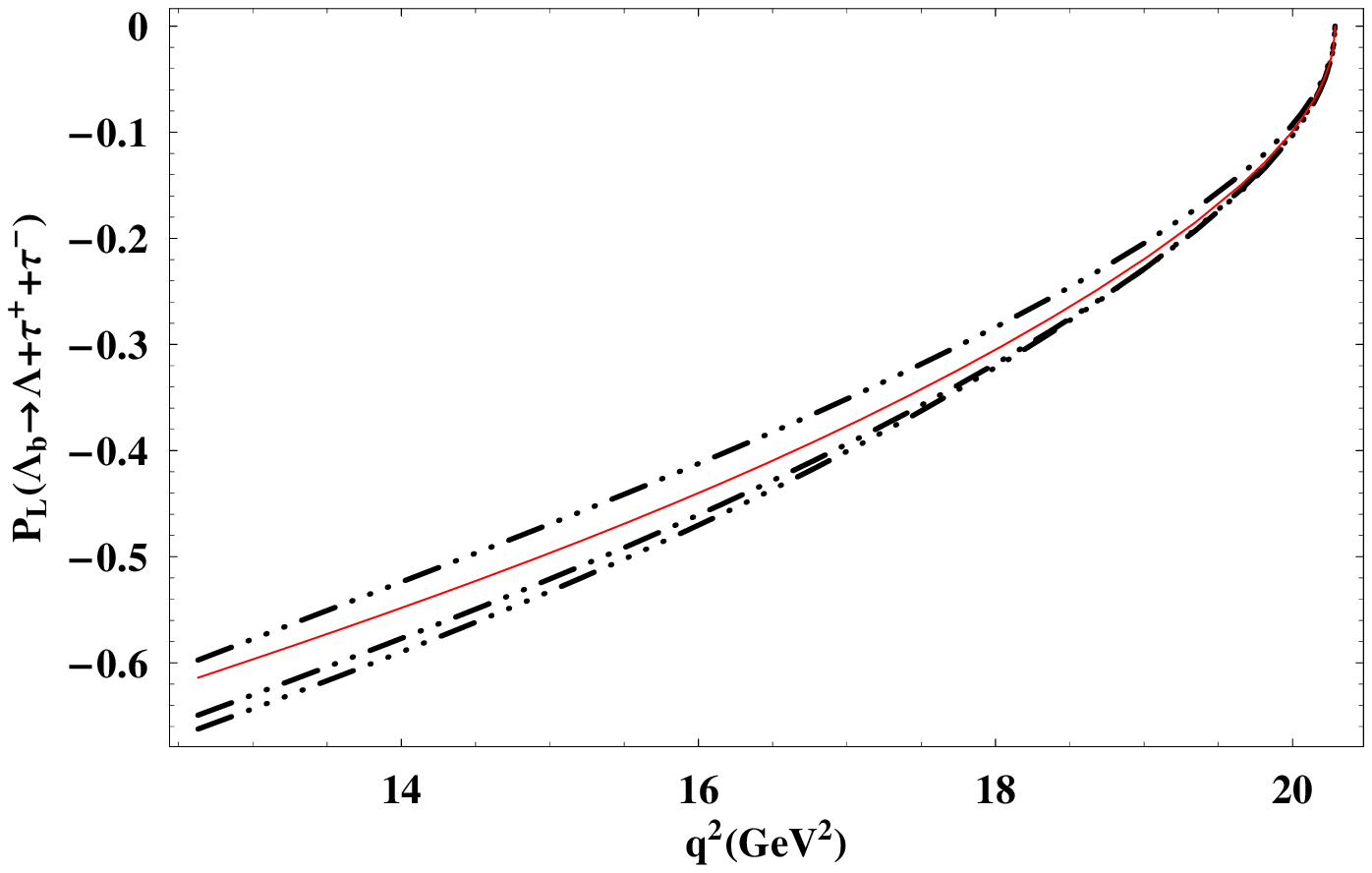}
\put (-350,290){(a)} \put (-100,290){(b)} \put (-350,0){(c)}
\put(-100,0){(d)}
\end{tabular}
\caption{(Color online). Longitudinal polarization asymmetry  of
$\Lambda_b \to \Lambda + l^{+} l^{-}$ ($l=\mu, \, \tau$) as a
function of momentum transfer $q^2$ at two fixed value of $1/R$ and
in the SM. The red line represents  the case in the SM;. The fig.
(a) and (b) describe the muon cases with the value of $1/R$ being
200 GeV and 500 GeV respectively; while  fig. (c) and (d)  reflect
the tauon cases with the value of $1/R$ being 200 GeV and 500 GeV.}
\label{longitudinal polarization asymmetry of semileptonic decay}
\end{center}
\end{figure}

\begin{figure}[tb]
\begin{center}
\begin{tabular}{ccc}
\includegraphics[scale=0.6]{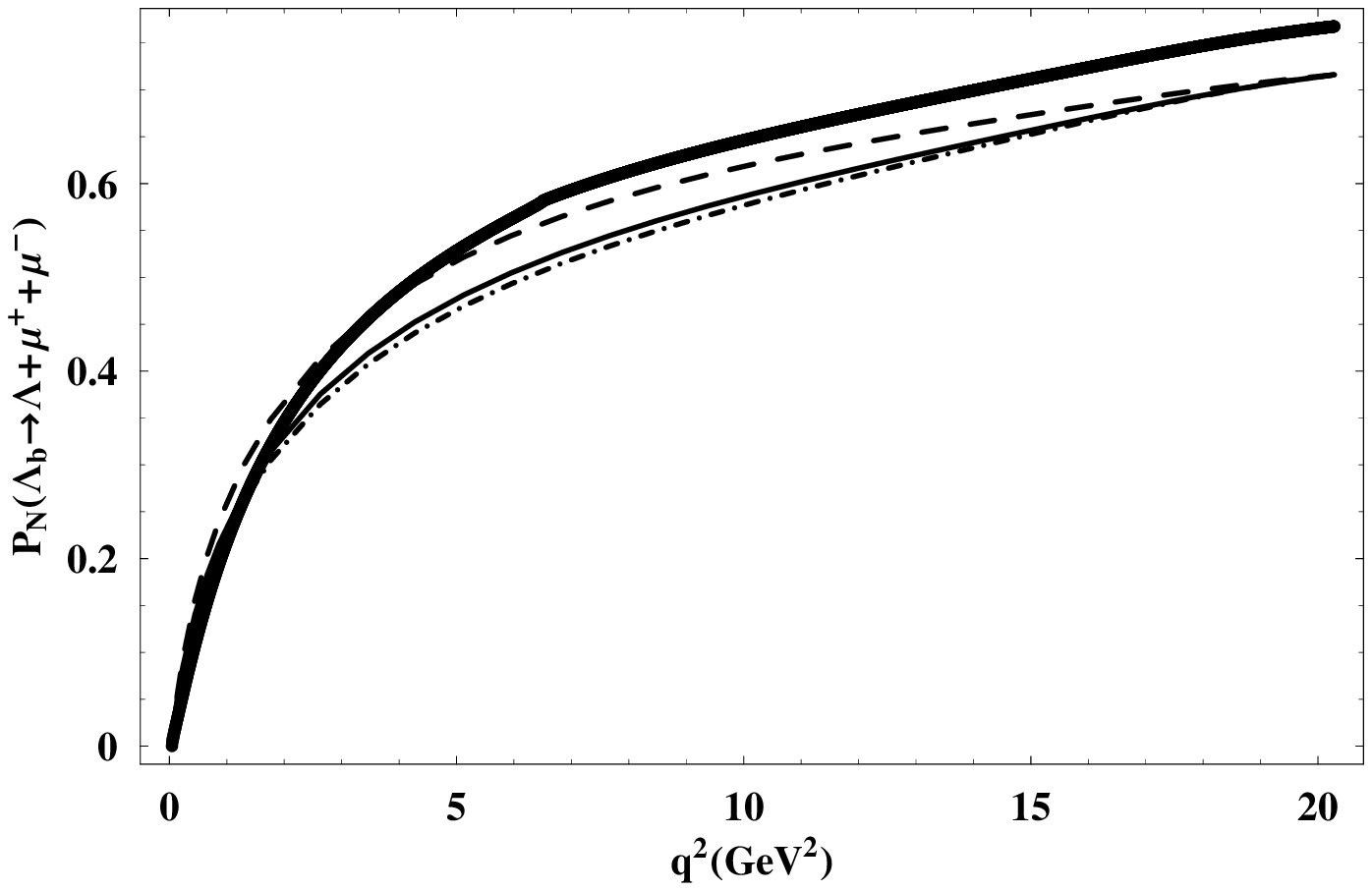}
\includegraphics[scale=0.6]{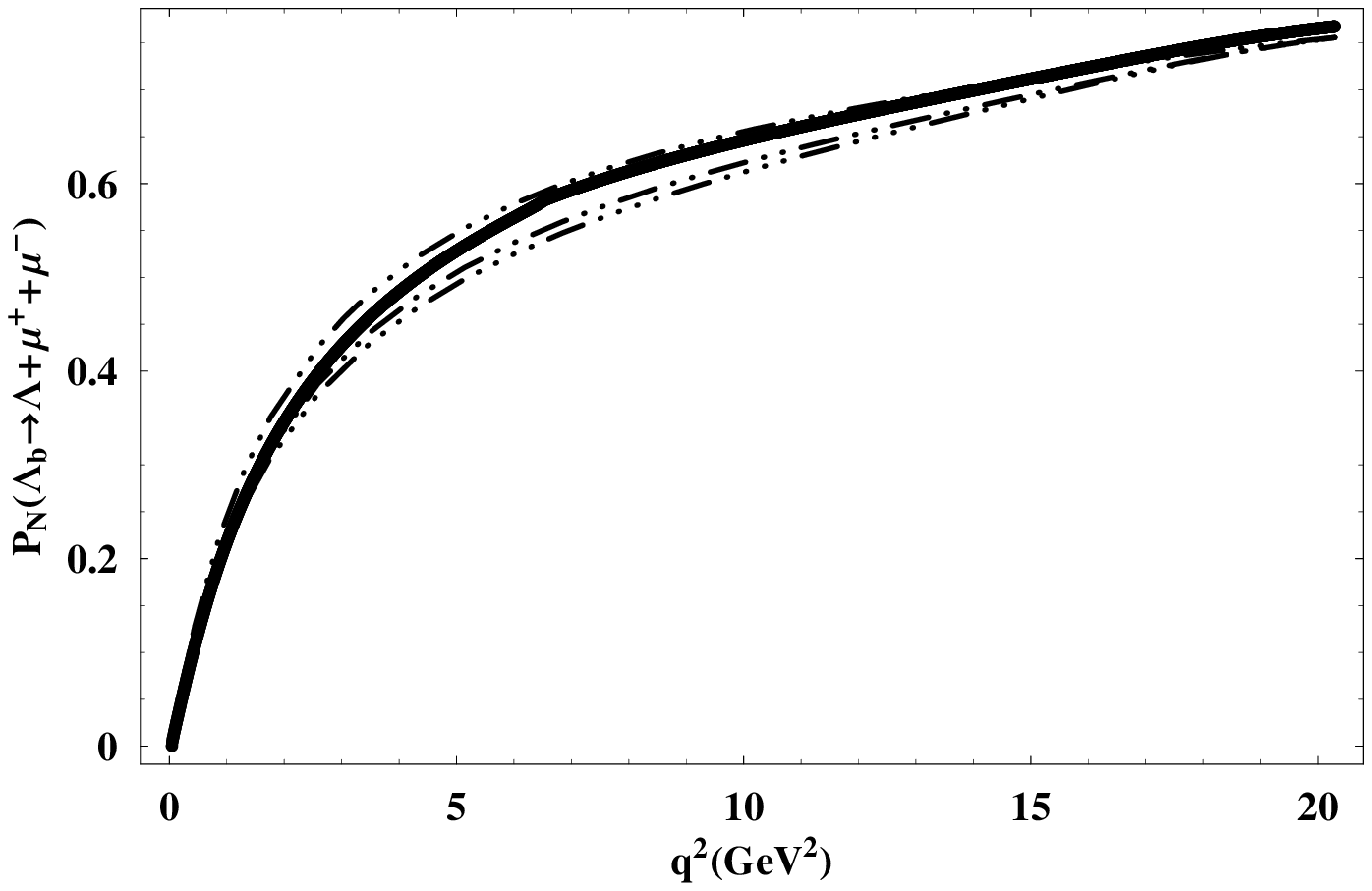}
\\
\includegraphics[scale=0.6]{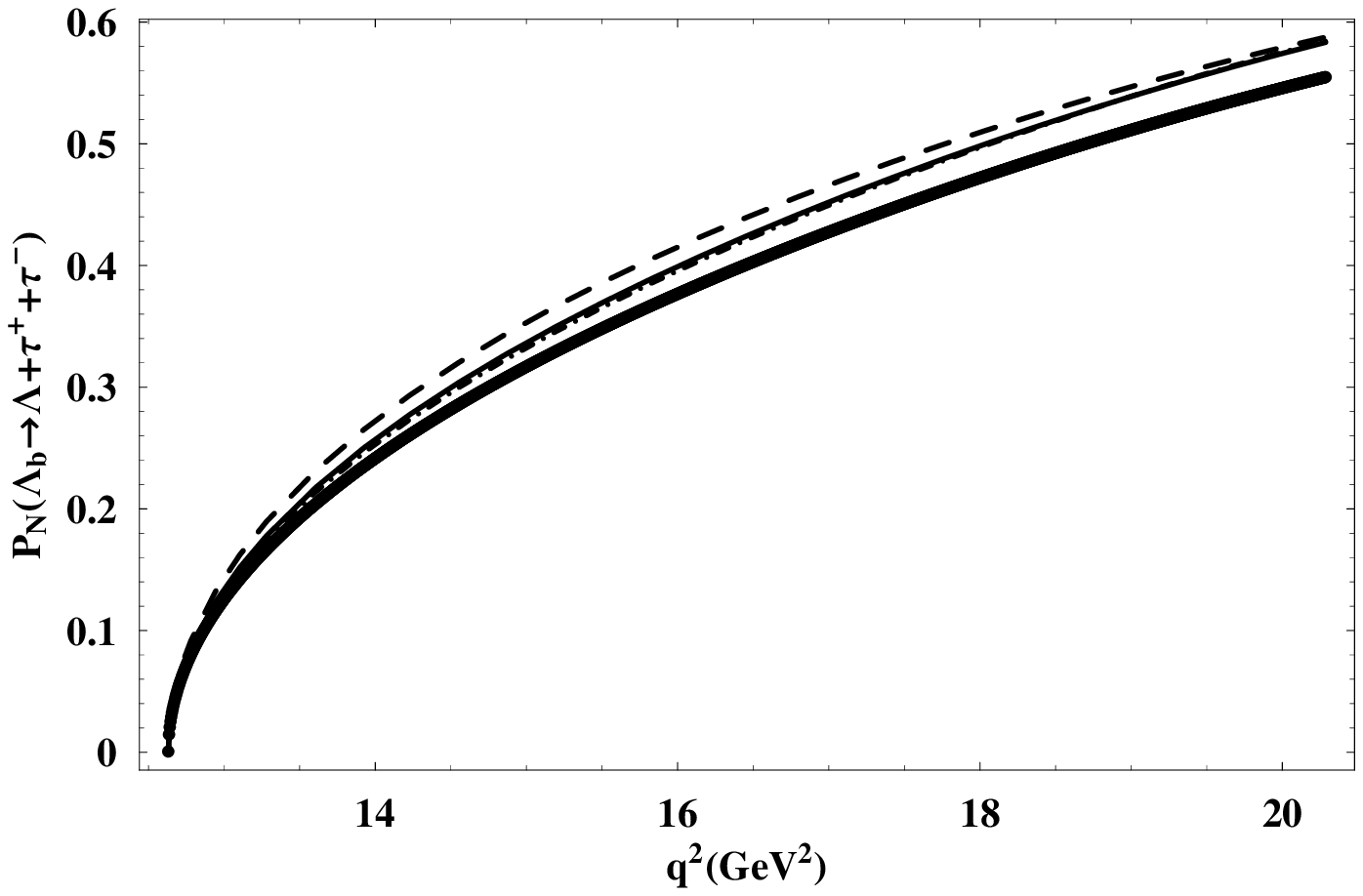}
\includegraphics[scale=0.6]{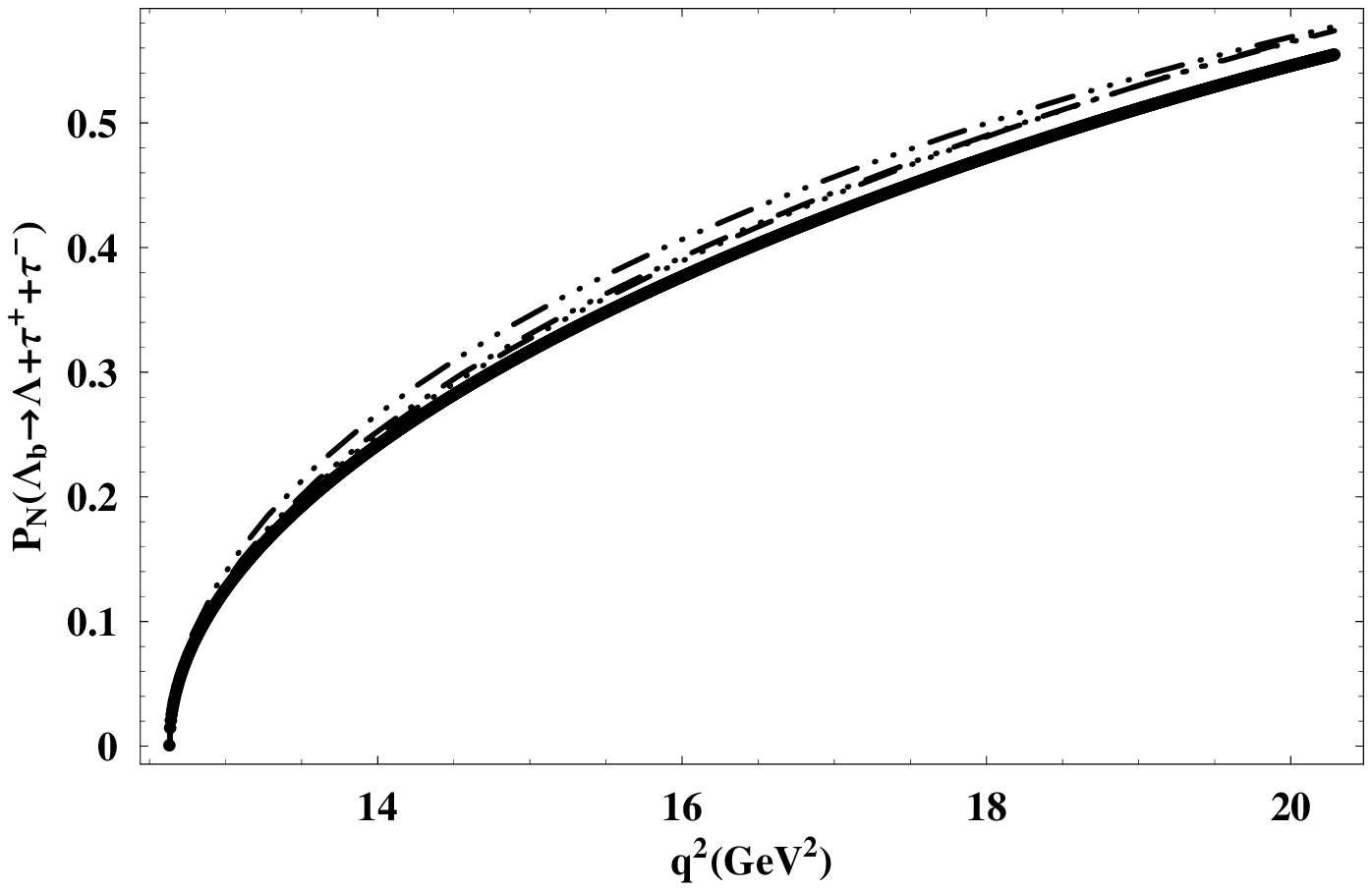}
\put (-350,290){(a)} \put (-100,290){(b)} \put (-350,0){(c)}
\put(-100,0){(d)}
\end{tabular}
\caption{Normal polarization asymmetry  of $\Lambda_b \to \Lambda +
l^{+} l^{-}$ ($l=\mu, \, \tau$) as a function of momentum transfer
$q^2$ at two fixed value of $1/R$ and in the SM. The bold line
represents the case in the SM;. The fig. (a) and (b) describe the
muon cases with the value of $1/R$ being 200 GeV and 500 GeV
respectively; while  fig. (c) and (d)  reflect the tauon cases with
the value of $1/R$ being 200 GeV and 500 GeV.} \label{normal
polarization asymmetry of semileptonic decay}
\end{center}
\end{figure}


\section{Conclusions}
In this paper, we investigate the exclusive weak decay of $\Lambda_b
\to \Lambda \gamma$ and $\Lambda_b \to \Lambda l^{+} l^{-}$ ($l=\mu
\, \tau$) in a single universal extra dimension scenario, which is a
strong contender to explore physics beyond the SM.  The priority to
investigate the bottom decays can be attributed to their sensitivity
of the flavor structure of nature, which leads to an extremely rich
phenomenology. More important, the large mass of heavy quark makes
the troublesome strong interaction effects controllable within heavy
quark expansion on the theoretical side, allowing for theoretical
predictions of acceptable accuracy. The form factors responsible for
$\Lambda_b \to \Lambda$ transition used in this paper are borrowed
from that calculated in the LCSR approach, where the higher twist
distribution amplitudes of $\Lambda$ baryon are included.

Due to the suppression of Wilson coefficient $C_7$ in the ACD model,
we find that the branching fraction $BR(\Lambda_b \to \Lambda
\gamma)$ is suppressed by  25\% for $1/R=300 {\rm{GeV}}$ compared
with that in the SM, which is similar to the inclusive $B \to X_s
\gamma$ decay \cite{Buras:2003mk}. However, the contributions from
KK modes can give rise to  10\% enhancement for fixed value of
$1/R=300 {\rm{GeV}}$ as a consequence of  larger number of Wilson
coefficient $C_{10}$ compared with that in the SM. Besides, it is
found that the zero-position of forward-backward asymmetry for
$\Lambda_b \to \Lambda \mu^{+} \mu^{-}$ is sensitive on the radius
of extra dimension $R$, which can be used to probe the new physics
effectively once the experimental data are available. The
longitudinal and transverse polarization asymmetries of $\Lambda$
baryon for the decays of  $\Lambda_b \to \Lambda l^{+} l^{-}$ are
found to be insensitive to the effect of extra dimension in the ACD
model. For the case of large momentum transverse $q^2$, the normal
polarization asymmetry $P_N(q^2)$ of  $\Lambda_b \to \Lambda l^{+}
l^{-}$ in the UED model can be marginally distinguishable from that
in the SM. Absolutely, it is also worth to extend the analysis of
$\Lambda_b \to \Lambda$ transition presented here to the case
 of $\Lambda_b$ decays to heavier $\Lambda$- baryons (resonance),
 which may be another interesting field to explore the effects from
extra dimensions and will be investigated in our future work.

\section*{Acknowledgements}

This work is partly supported by National Science Foundation of
China under Grant No.10735080 and 10625525. The authors would like
to thank Cheng Li and Yue-Long Shen for helpful discussions.


\begin{thebibliography}{99}


\bibitem{HFAG}The Heavy Flavor Averaging Group, arXiv:0808.1297 [hep-ex], and
online update at http://www.slac.stanford.edu/xorg/hfag/ .



\bibitem{Mannel}T.~Mannel and S.~Recksiegel,
  J.\ Phys.\ G {\bf 24} (1998) 979
  [arXiv:hep-ph/9701399].



\bibitem{Mohanta} R.~Mohanta, A.~K.~Giri, M.~P.~Khanna, M.~Ishida and S.~Ishida,
  Prog.\ Theor.\ Phys.\  {\bf 102} (1999) 645
  [arXiv:hep-ph/9908291].




\bibitem{Cheng} H.~Y.~Cheng, C.~Y.~Cheung, G.~L.~Lin, Y.~C.~Lin, T.~M.~Yan and H.~L.~Yu,
  Phys.\ Rev.\  D {\bf 51} (1995) 1199
  [arXiv:hep-ph/9407303].



\bibitem{Cheng 2}H.~Y.~Cheng and B.~Tseng,
  Phys.\ Rev.\  D {\bf 53}, 1457 (1996)
  [Erratum-ibid.\  D {\bf 55}, 1697 (1997)]
  [arXiv:hep-ph/9502391].


\bibitem{Huang}C.~S.~Huang and H.~G.~Yan,
  Phys.\ Rev.\  D {\bf 59} (1999) 114022
  [Erratum-ibid.\  D {\bf 61} (2000) 039901]
  [arXiv:hep-ph/9811303].


\bibitem{Hiller:2001zj}
  G.~Hiller and A.~Kagan,
  Phys.\ Rev.\  D {\bf 65} (2002) 074038
  [arXiv:hep-ph/0108074].


\bibitem{HLLW}X.~G.~He, T.~Li, X.~Q.~Li and Y.~M.~Wang,
  Phys.\ Rev.\  D {\bf 74} (2006) 034026
  [arXiv:hep-ph/0606025].



\bibitem{Aslam:2008hp}
  M.~J.~Aslam, Y.~M.~Wang and C.~D.~Lu,
  arXiv:0808.2113 [hep-ph].


\bibitem{Antoniadis:1990ew}
  I.~Antoniadis,
  Phys.\ Lett.\  B {\bf 246} (1990) 377.

  

\bibitem{ACD}T.~Appelquist, H.~C.~Cheng and B.~A.~Dobrescu,
  Phys.\ Rev.\  D {\bf 64} (2001) 035002
  [arXiv:hep-ph/0012100].



\bibitem{Appelquist:2002wb}
  T.~Appelquist and H.~U.~Yee,
  Phys.\ Rev.\  D {\bf 67} (2003) 055002
  [arXiv:hep-ph/0211023].



\bibitem{Agashe:2001ra}
  K.~Agashe, N.~G.~Deshpande and G.~H.~Wu,
  Phys.\ Lett.\  B {\bf 511} (2001) 85
  [arXiv:hep-ph/0103235].




\bibitem{Appelquist:2001jz}
  T.~Appelquist and B.~A.~Dobrescu,
  Phys.\ Lett.\  B {\bf 516} (2001) 85
  [arXiv:hep-ph/0106140].



\bibitem{Oliver:2002up}
  J.~F.~Oliver, J.~Papavassiliou and A.~Santamaria,
  Phys.\ Rev.\  D {\bf 67} (2003) 056002
  [arXiv:hep-ph/0212391].











\bibitem{Buras:2002ej}
  A.~J.~Buras, M.~Spranger and A.~Weiler,
  Nucl.\ Phys.\  B {\bf 660} (2003) 225
  [arXiv:hep-ph/0212143].



\bibitem{Buras:2003mk}
  A.~J.~Buras, A.~Poschenrieder, M.~Spranger and A.~Weiler,
  Nucl.\ Phys.\  B {\bf 678} (2004) 455
  [arXiv:hep-ph/0306158].



\bibitem{Haisch:2007vb}
  U.~Haisch and A.~Weiler,
  Phys.\ Rev.\  D {\bf 76} (2007) 034014
  [arXiv:hep-ph/0703064].




\bibitem{Colangelo}  P.~Colangelo, F.~De Fazio, R.~Ferrandes and T.~N.~Pham,
  Phys.\ Rev.\  D {\bf 73} (2006) 115006
  [arXiv:hep-ph/0604029]; I.~Ahmed, M.~A.~Paracha and M.~J.~Aslam,
  Eur.\ Phys.\ J.\  C {\bf 54} (2008) 591
  [arXiv:0802.0740 [hep-ph]];
  A.~Saddique, M.~J.~Aslam and C.~D.~Lu,
  Eur.\ Phys.\ J.\  C {\bf 56} (2008) 267
  [arXiv:0803.0192 [hep-ph]].



\bibitem{Mohanta:2006ae}
  R.~Mohanta and A.~K.~Giri,
  Phys.\ Rev.\  D {\bf 75} (2007) 035008
  [arXiv:hep-ph/0611068].


\bibitem{Colangelo:2007jy}
  P.~Colangelo, F.~De Fazio, R.~Ferrandes and T.~N.~Pham,
  Phys.\ Rev.\  D {\bf 77} (2008) 055019
  [arXiv:0709.2817 [hep-ph]].






\bibitem{Aliev:2006xd}
  T.~M.~Aliev and M.~Savci,
  Eur.\ Phys.\ J.\  C {\bf 50} (2007) 91
  [arXiv:hep-ph/0606225].



\bibitem{Aliev:2006gv}
  T.~M.~Aliev, M.~Savci and B.~B.~Sirvanli,
  Eur.\ Phys.\ J.\  C {\bf 52} (2007) 375
  [arXiv:hep-ph/0608143].



\bibitem{Wang:2008sm}
  Y.~M.~Wang, Y.~Li and C.~D.~Lu,
  arXiv:0804.0648 [hep-ph].



\bibitem{Buchalla:1995vs}
  G.~Buchalla, A.~J.~Buras and M.~E.~Lautenbacher,
  Rev.\ Mod.\ Phys.\  {\bf 68} (1996) 1125
  [arXiv:hep-ph/9512380].


\bibitem{Buras:1993xp}
  A.~J.~Buras, M.~Misiak, M.~Munz and S.~Pokorski,
  Nucl.\ Phys.\  B {\bf 424} (1994) 374
  [arXiv:hep-ph/9311345].


\bibitem{Misiak:1992bc}
  M.~Misiak,
  Nucl.\ Phys.\  B {\bf 393} (1993) 23
  [Erratum-ibid.\  B {\bf 439} (1995) 461].



\bibitem{Buras:1994dj}
  A.~J.~Buras and M.~Munz,
  Phys.\ Rev.\  D {\bf 52} (1995) 186
  [arXiv:hep-ph/9501281].




\bibitem{c.q. geng 4}C.~H.~Chen and C.~Q.~Geng,
  Phys.\ Rev.\  D {\bf 64} (2001) 074001
  [arXiv:hep-ph/0106193].



  \bibitem{Aliev 1} T.~M.~Aliev, A.~Ozpineci and M.~Savci,
  Nucl.\ Phys.\  B {\bf 649} (2003) 168
  [arXiv:hep-ph/0202120].

\bibitem{Aliev 2} T.~M.~Aliev, A.~Ozpineci and M.~Savci,
  Phys.\ Rev.\  D {\bf 67} (2003) 035007
  [arXiv:hep-ph/0211447].

\bibitem{Aliev 3} T.~M.~Aliev, V.~Bashiry and M.~Savci,
  Nucl.\ Phys.\  B {\bf 709} (2005) 115
  [arXiv:hep-ph/0407217].

\bibitem{Aliev 4} T.~M.~Aliev and M.~Savci,
  JHEP {\bf 0605} (2006) 001
  [arXiv:hep-ph/0507324].






\bibitem{PDG}W.M. Yao {\it et al.,} J. Phys. G {\bf{ 33}}, 1 (2006).


\bibitem{b to s in theory 9}A. Ali,  T. Mannel and T. Morozumi, Phys. Lett.  B{\bf 273} (1991) 505.


\bibitem{c.q. geng 2} C.~H.~Chen and C.~Q.~Geng,
Phys.\ Rev.\ D \textbf{63} (2001) 114024 [arXiv:hep-ph/0101171].



\end{thebibliography}
\end{document}